\documentclass[10pt,final,doublecolumn]{IEEEtran}
\usepackage{subfigure}
\usepackage{amsmath}
\usepackage{latexsym}
\usepackage{graphicx}
\usepackage{bbding}
\usepackage{indentfirst}
\usepackage{setspace}
\usepackage{cite,balance}
\usepackage{float}
\usepackage{epstopdf}
\usepackage{amssymb}
\usepackage{amsfonts}
\usepackage{enumerate}
\usepackage{multicol}
\usepackage{color}
\usepackage{stfloats}
\usepackage{bm}
\usepackage{amssymb}
\usepackage{stfloats}
\usepackage{algorithm,algpseudocode,float}
\usepackage{lipsum}

\newtheorem{remark}{\bf \emph{\underline{Remark}}}

\def\({\left(}
\def\){\right)}

\def\b0{{\mathbf{0}}}

\IEEEoverridecommandlockouts
\allowdisplaybreaks[4]

\makeatletter
\newenvironment{breakablealgorithm}
  {
   \begin{center}
     \refstepcounter{algorithm}
     \hrule height.8pt depth0pt \kern2pt
     \renewcommand{\caption}[2][\relax]{
       {\raggedright\textbf{\ALG@name~\thealgorithm} ##2\par}%
       \ifx\relax##1\relax 
         \addcontentsline{loa}{algorithm}{\protect\numberline{\thealgorithm}##2}%
       \else 
         \addcontentsline{loa}{algorithm}{\protect\numberline{\thealgorithm}##1}%
       \fi
       \kern2pt\hrule\kern2pt
     }
  }{
     \kern2pt\hrule\relax
   \end{center}
  }
\makeatother

\begin{document}

\title{Reconfigurable Intelligent Surface-Aided 6G Massive Access: Coupled Tensor Modeling and Sparse Bayesian Learning}
\author{{Xiaodan Shao,~\IEEEmembership{Member,~IEEE}, Lei Cheng,
Xiaoming Chen, \IEEEmembership{Senior Member, IEEE}, Chongwen Huang, \IEEEmembership{Member, IEEE}, and Derrick Wing Kwan Ng, \IEEEmembership{Fellow, IEEE}}
    \vspace{-18pt}
	\thanks{Xiaodan Shao is with the College of Information Science and Electronic Engineering, Zhejiang University, Hangzhou, 310027, China, and also with  Zhejiang Provincial Key Laboratory of Info. Proc., Commun. \& Netw. (IPCAN), Hangzhou 310027, China (Email: {\tt shaoxiaodan@zju.edu.cn}).}

\thanks{Lei Cheng is with the College of Information Science and Electronic Engineering, Zhejiang University, Hangzhou, 310027, China, and also with  Zhejiang Provincial Key Laboratory of Info. Proc., Commun. \& Netw. (IPCAN), Hangzhou 310027, China (e-mail: {\tt lei\_cheng@zju.edu.cn}).}

\thanks{Xiaoming Chen is with the College of Information Science and Electronic Engineering, Zhejiang University, Hangzhou 310027, China (Email: {\tt chen\_xiaoming@zju.edu.cn}).}

\thanks{C. Huang is with College of Information Science and Electronic Engineering, Zhejiang University, Hangzhou 310027, China, and with International Joint Innovation Center, Zhejiang University, Haining 314400, China, and also with Zhejiang-Singapore Innovation and AI Joint Research Lab and Zhejiang Provincial Key Laboratory of Info. Proc., Commun. \& Netw. (IPCAN), Hangzhou 310027, China. (E-mail: {\tt chongwenhuang@zju.edu.cn}).}

\thanks{Derrick Wing Kwan Ng  is with the School of Electrical Engineering and Telecommunications, University of New South Wales, Sydney, NSW 2052, Australia.(E-mail: {\tt w.k.ng@unsw.edu.au}).}
}
\maketitle

\begin{abstract}
This paper investigates a reconfigurable intelligent surface (RIS)-aided unsourced random access (URA) scheme
for the sixth-generation (6G) wireless networks with massive sporadic traffic devices. First of all, this paper proposes a novel joint active device separation (the message recovery of active device) and channel estimation architecture for the RIS-aided URA. Specifically, the RIS passive reflection is optimized before the successful device separation. Then, by associating the data sequences to multiple rank-one tensors and exploiting the angular sparsity of the RIS-BS channel, the detection problem is cast as a high-order coupled tensor decomposition problem without the need of exploiting pilot sequences. However, the inherent coupling among multiple sparse device-RIS channels, together with the unknown number of active devices make the detection problem at hand deviate from the widely-used coupled tensor decomposition format. To overcome this challenge, this paper judiciously devises a probabilistic model that captures both the element-wise sparsity from the angular channel model and the low-rank property due to the sporadic nature of URA. Then, based on such a probabilistic model, a iterative detection algorithm is developed under the framework of sparse variational inference, where each update iteration is obtained in a closed-form and the number of active devices can be automatically estimated for effectively avoiding the overfitting of noise. Extensive simulation results confirm the excellence of the proposed URA algorithm, especially for the case of a large number of reflecting elements for accommodating a significantly large number of devices.
\end{abstract}


\IEEEpeerreviewmaketitle

\section{Introduction}
Massive access or massive machine-type communication (mMTC) has been identified as one of the main application cases in the upcoming sixth-generation (6G) wireless networks, where the number of potential devices is expected to reach hundreds of billions by 2030. In particular, a large number of devices are connected to the Internet via a base station (BS) with sporadic communications, where only a small fraction of devices are active concurrently \cite{coop}. In this context, applying conventional grant-based random access schemes to this emerging scenario would lead to an exceedingly high access latency and a prohibitive signaling overhead. To address this issue, grant-free random access schemes have been regarded as a promising candidate technology for enabling massive access in 6G wireless networks \cite{6G}, where active devices transmit their data without establishing any access grant protocol with the BS. Currently, a commonly discussed category of grant-free random access is sourced random access, where each device is preassigned with a unique non-orthogonal pilot sequence and the BS identifies the device activity based on the received signals by distinguishing the transmitted sequences \cite{coop}. However, since the number of devices in 6G wireless network grows sufficiently large, sourced random access schemes become increasingly inefficient to assign fixed pilot sequences to all potential devices. For instance, in order to detect $K_a$ active devices from total $\bar{K}$ devices, the length of pilot sequence should grow with the scale of $K_a\log(\bar{K})$, even through an effective sparse recovery algorithm implemented \cite{ampgao,shaoiot}.

To overcome this challenge, another category of grant-free random access, i.e., unsourced random access (URA), was proposed in \cite{PersUra}.
Indeed, the roll-out of URA is motivated by practical Internet-of-Things scenarios, where millions of low-cost devices have a common codebook at the moment of production \cite{PersUra}. Active devices maps their messages to codewords of the common codebook, and the BS recovers the messages by detecting the transmitted codewords, without determining the identities of active devices. If active devices wish to reveal themselves, they can embed identity information in their payloads. In order to decrease the size of codebook, active devices usually partition their messages to several sub-messages \cite{unsourced,bayf}. Active devices adds some parity check bits to each sub-message which is mapped to a codeword of the common codebook, and then the BS recovers all the sub-messages which are stitched to the original messages by using a some decoder according to the parity check bits, e.g., the tree-based decoder \cite{unsourced}. Compared to sourced random access, URA has a higher spectral efficiency, since it does send pilot sequence for identity detection \cite{wuyongUra}. Moreover, URA has a lower computational complexity, which is independent of the total number of potential devices and depends only on the cardinality of the active device set \cite{wuyongUra}. Hence, URA is appealing to 6G wireless networks with a large number of devices with sporadic activities.

Despite the advancement of URA, the devices may locate in a service dead zone, where the direct link from the devices to the BS is not available \cite{shi}. Consequently, the signals received from the devices are weak such that performing active device separation is a challenging task for the BS. To overcome this challenge, the use of millimeter-wave (mmWave)/terahertz (THz) frequencies with massive multiple-input-multiple output (MIMO) was proposed to enhance the detection accuracy of grant-free random access \cite{shaommv}. However, it significantly increases the hardware and energy cost as well as signal processing complexity due to the deployment of a large number of radio frequency (RF) chains. To this end, emerging wireless technologies for manipulating the wireless channels with low cost, low complexity, and high energy efficiency are desired to unlock the potential of URA in 6G wireless networks.

Recently, reconfigurable intelligent surface (RIS), as a promising technology of 6G wireless networks, has been proposed to enhance the signal quality at desired receivers \cite{kwanro, rui, pan,imag, huang,irssensing}. Specifically, an RIS is able to establish favourable channel responses by customizing the wireless propagation environment via a large number of reconfigurable passive reflecting elements. Thus, the deployment of RIS can enhance the strength of the signals between the BS and the devices in dead zones via bypassing the obstacle between them. Moreover, by optimizing the phase shifts, the signal propagation between the BS and the devices can be smartly reconfigured to improve the accuracy of active device separation.
On the other hand, the performance gain in RIS-aided communication systems depends critically on the availability of channel state information (CSI). Yet, CSI acquisition is quite challenging in practice due to the passive nature of RIS. In general, RIS without equipping active radio frequency chains can neither transmit nor receive pilot signals, thus it is challenging to estimate the channels from RIS to the BS
and devices to RIS separately. Instead, the concatenated device-RIS-BS channels are usually estimated based on the pilot sequences sent from the devices. For example, by exploiting the information on the slow-varying channel components and the hidden channel sparsity, \cite{RISsparse} formulated the channel estimation problem for RIS-aided multiuser MIMO systems as a matrix-calibration based matrix factorization task. Besides, to enable massive random access, \cite{shi} proposed a three-stage framework based on the approximate message passing (AMP) for joint active device separation and channel estimation in RIS-aided sourced random access.

Although various approaches have been proposed in the literature for channel estimation in RIS-aided systems, e.g., \cite{rui,pan,huang,RISsparse,shi},
the required pilot signaling overhead scales with the product of the number of RIS reflecting elements and devices, which are prohibitively large in practical scenarios, especially for the case of massive access. As a remedy, the study of joint active device separation and channel estimation with a fewer number of pilots for the RIS-aided URA scheme in 6G wireless networks is desired. To achieve the goal, the considered problem can be formulated as a novel non-standard coupled tensor decomposition problem with a sparse coupled factor, assuming no knowledge about the tensor rank (i.e., the number of active devices). To handle coupled tensor decomposition type problems, various attempts have been conducted \cite{cou,cou2}. For instance, \cite{cou} proposed a coupled polyadic decomposition method, where each tensor is a sum of coupled rank-1 tensors sharing some common factors.  Unfortunately, the results from \cite{cou} and \cite{cou2} cannot be directly applied to the joint active device separation and channel estimation problem with a complicated coupled structure.
Furthermore, the number of active devices is unknown in practice, which determines the number of unknown model parameters. Besides, it is generally non-deterministic polynomial-time hard (NP-hard) to acquire the number of active devices from tensor data. To address this problem, the authors in \cite{tenrank} introduced an additive
regularization term that penalizes complicated channel models to mitigate the overfitting of noise. However, this method inevitably consumes enormous computational resources and leads to a heavy computation burden. To learn the tensor rank automatically, recent advances integrate the tensor rank learning into its hyper-parameter inference steps, such that the Bayesian theory can provide a natural recipe for automatic rank determination \cite{chengpro, zhao, cheng1, cheng2}. Nevertheless, the effectiveness of applying variational Bayesian inference for solving coupled tensor decomposition is still unclear in the literature. Moreover, the coupling structure (from the RIS-aided URA scheme \cite{PersUra}), the element-wise sparsity structure (from angular channel representation \cite{power}), and the low-rank property (from the sporadic nature of mMTC \cite{6G}) together pose unique challenges in designing proper probabilistic model and the associated inference algorithm. In this paper, by devising novel priors to incorporate problem structures, we design a novel probabilistic model to investigate the trade-off between the expressive power of the model and the tractability of the inference algorithm. The contributions of this paper are as follows:
\begin{enumerate}
\item
    This paper studies a new application of RIS for detecting active devices and estimating the corresponding channels to support URA in 6G wireless networks. Specifically, we propose a new RIS-aided URA architecture, where RIS passive reflection is first optimized for enhancing the performance of device separation. Then the transmitted data symbols are constructed as multiple rank-1 tensors that allows a pragmatic method for active device separation and device-RIS channel estimation without relying on pilot sequences.

\item
    This paper proposes a novel element-wise sparsity and low-rank inducing probabilistic model for the considered high-order coupled tensor decomposition problem. By using variational inference, an intelligent joint active device separation and channel estimation algorithm is derived, assuming no knowledge of active device number and noise power.

\item This paper provides a thorough theoretical analysis of the proposed algorithm and the results show that it enjoys fast convergence and low computational complexity. Importantly, the proposed algorithm is well designed for handling URA in large-scale RIS regime compared with the existing ones.
\end{enumerate}

The rest of this paper is organized as follows: Section II introduces the system model and the channel models, and Section III proposes the RIS-aided joint active device separation and channel estimation architecture. Section IV designs a coupled tensor based automatic detection algorithm by applying Bayesian learning and analyzes the performance of the proposed algorithm. Section V provides extensive simulation results to validate the effectiveness of the proposed algorithm. Finally, Section VI concludes the paper.

\emph{Notations}: We use $\mathbb{C}^{A\times B}$ to denote the space of complex matrices of size $A\times B$, $|\cdot|$ to denote the absolute value of a complex number, $(\cdot)^H$ and $(\cdot)^T$ to denote conjugate transpose and transpose, respectively, $\otimes$ to denote the Kronecker product, $\circ$ to denote vector outer product, $\left\|\cdot\right\|_F$ to denote Frobenius norm of a matrix, $\left\|\cdot\right\|_2$ to denote the
$l_2$-norm of an input vector, $\left[\!\left[\cdot\right]\!\right]$ to denote the Kruskal operator. $x\in [a,b]$ means that scalar $x$ lies in the closed interval between $a$ and $b$. $\mathbb{E}$ denotes the expectation of its input argument. $\|\cdot\|_{0}$ denotes the $l_{0}$-norm defined as the number of nonzero elements of an input vector or matrix. $*$ denotes conjugation. The Khatri-Rao product is denoted by $\diamond$. The Hadamard product is denoted by $\odot$. $\mathbf{I}_{K}$ represents the $K\times K$ identity matrix. Vector $\mathbf{1}_K \in \mathbb{R}^K$ is the all-ones vector with length $K$. $\mathbf{a}(n)$ denotes the $n$th element of vector $\mathbf{a}$. $\mathbf{A}(n,k)$ denotes the $(n,k)$th element of matrix $\mathbf{A}$. $\mathbf{A}(:,n)$ and $\mathbf{A}(n,:)$ denote the $n$th column and the $n$th row of matrix $\mathbf{A}$, respectively. $\mathrm{Tr}(\cdot)$ denotes the trace of a matrix. $\mathbf{x}\in \mathbb{C}^{n}$ is said to follow a vector-valued normal distribution with mean $\mathbf{u}$ and covariance matrix $\boldsymbol{\Sigma}$ of the form $\mathcal{CN}_{n}(\mathbf{x}|\mathbf{u},\boldsymbol{\Sigma})$.
Matrix $\mathbf{G}$ is said to follow the matrix-variate normal distribution with mean matrix $\mathbf{M}$ and covariance matrix $\boldsymbol{\Upsilon}\otimes\boldsymbol{\Sigma}$ of the form $\mathcal{CN}_{N\times K}(\mathbf{G}|\mathbf{M},\boldsymbol{\Upsilon}\otimes\boldsymbol{\Sigma})=\frac{1}{(2\pi)^{NK/2}|\boldsymbol{\Upsilon  }|_d^{N/2}|\boldsymbol{\Sigma}|_d^{K/2}}\exp(-\frac{1}{2}\mathrm{Tr}\boldsymbol{\Upsilon }^{-1}(\mathbf{G}-\mathbf{M})\boldsymbol{\Sigma}^{-1}(\mathbf{G}-\mathbf{M})^H)$ with $|\cdot|_d$ being the determinant. $\mathcal{O}(\cdot)$ stands for the big-O notation. $p(\cdot|\cdot)$ denotes conditional probability distribution. $\mathrm{diag}(\mathbf{x})$ denotes a diagonal matrix with the diagonal entries specified by vector $\mathbf{x}$. $\mathcal{D}[\mathbf{A}]$ is a diagonal matrix taking the diagonal element from $\mathbf{A}$.

\section{System Model}
\begin{figure}[t]
  \centering
\includegraphics [width=85mm] {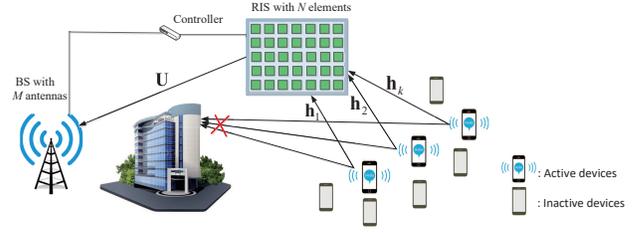}
\caption{RIS-aided massive unsourced random access in 6G wireless networks with blocked direct links.}
\label{sys}
\end{figure}

We consider a RIS-aided 6G wireless network, as illustrated in Fig. \ref{sys}, where a BS equipped with $M$ antennas serves massive single-antenna devices. Herein, the RIS consisting of $N$ passive reflecting elements is deployed to enhance the devices' communication performance. The passive reflecting elements of the RIS are arranged as
an $N_1\times N_2$ uniform rectangular array with $N\triangleq N_1\times N_2$. Although the number of potential devices $\bar{K}$ is numerous, only $K_a \ll \bar{K}$ devices are active concurrently at a given slot, due to the bursty nature of the envisioned 6G services \cite{6G}. Note that the number of active devices, $K_a$, is random and unknown to the BS. It is assumed that the direct links between active devices and the BS are obstructed \cite{shi}, and thus calls for a need to enhance the quality of communications via deploying RIS \cite{rui}. The channels are assumed to be invariant in a given time slot, but might independently fade over time slots. In particular, $\mathbf{h}_k\in \mathbb{C}^{N}$ is used to denote the channel vector between the RIS and the $k$th device; $\mathbf{U}\in \mathbb{C}^{M\times N}$ is used to denote the channel matrix between the BS and the RIS.

\subsection{Channel Models}
Since the RIS is usually mounted at some tall building, it is expected that there are only limited scatters around the RIS. This suggests the adoption of a sparse angular channel model \cite{power}, which has been validated by real-world measurements and widely adopted in RIS related research \cite{RISsparse}. In specific,
the RIS-BS channel, i.e., ${\bf U}\in \mathbb{C}^{M\times N}$ can be expressed
as
\begin{align}\label{U}
\mathbf{U}=\sqrt{\bar{\mu}}\sum_{i=1}^{\bar{I}}\bar{\xi}_i\mathbf{a}_\mathrm{B}(\sigma_i)\mathbf{a}_\mathrm{R}(\bar{\phi}_i,\bar{\psi}_i)^H,
\end{align}
where $
\mathbf{a}_\mathrm{R}(\bar{\phi}_i,\bar{\psi}_i)=\boldsymbol{\varphi}_{N_2}(\cos(\bar{\psi}_i)\cos(\bar{\phi}_i))\otimes \boldsymbol{\varphi}_{N_1}(\cos(\bar{\psi}_i)\sin(\bar{\phi}_i))$
with
$
\boldsymbol{\varphi}_{N_p}(x)\triangleq \frac{1}{\sqrt{N_p}}[1, \\ e^{-j\frac{2\pi}{\varrho}dx}, \cdots,e^{-j\frac{2\pi}{\varrho}d(N_p-1)x}]^T, \forall p=1,2
$; $(\bar{\phi}_i,\bar{\psi}_i)$ being the azimuth and elevation angle-of-departure (AoD) from the RIS, respectively; $\mathbf{a}_\mathrm{B}(\sigma_i)=\boldsymbol{\varphi}_{M}(\sin(\sigma_i))$ is the receive
array response with respect to AoDs from the RIS to the BS with $\|\mathbf{a}_\mathrm{B}(\sigma_i)\|^2=M$;
$\sigma_i$ is the corresponding azimuth angle-of-arrival
(AoA) to the BS;
$N_1$ and $N_2$ denote the length and the width of the rectangular array of RIS; $\varrho$ denotes the wavelength of carrier frequency;
$\bar{I}$ is the total number of spatial paths between the BS and the RIS; $\bar{\mu}$ is the RIS-BS distance-dependent path loss; $\bar{\xi}_i$ denotes the complex amplitude associated with the $i$-th path.

For the device-RIS channel $\mathbf{h}_k\in \mathbb{C}^{N}$, it can be modeled as
\begin{align}\label{angu}
\mathbf{h}_k=\sqrt{\mu_k}\sum_{i=1}^{I_k}\epsilon_i^k\mathbf{a}_\mathrm{R}(\phi_i^k,\psi_i^k),
\end{align}
where {$\epsilon_i^k$ denotes the complex amplitude associated with the $i$-th path; $\mu_k$ is the device-RIS distance-dependent path loss}; $\phi_i^k$ and $\psi_i^k$ are the
corresponding azimuth and elevation AoA to
the RIS, respectively; $I_k$ denotes the number of paths between the $k$-th device and the RIS.
Following the grid-based scheme in \cite{RISsparse}, the representation of channel vector $\mathbf{h}_k$ can be further simplified. In particular, two sampling grids $\boldsymbol{\nu}=[\nu_1,\cdots,\nu_{{N_1}^{'}}]^T$ with length ${N_1}^{'}\geq N_1$,
and $\boldsymbol{\varsigma }=[\varsigma_1,\cdots,\varsigma_{{N_1}^{'}}]^T$ with length ${N_2}^{'}\geq N_2$, are employed such that $\mathbf{h}_k$ can be represented as
\begin{align}\label{angu3}
&\mathbf{h}_k=\mathbf{A}_{\mathrm{R}}\boldsymbol{\lambda}_k,
\end{align}
where
$
\mathbf{A}_{\mathrm{R}}\! =\! \left[\boldsymbol{\varphi}_{N_1}(\nu_1),\cdots,\boldsymbol{\varphi}_{N_1}(\nu_{{N_1}^{'}})\right]\!\otimes \! \left[\boldsymbol{\varphi}_{N_2}(\varsigma_1),\cdots,\boldsymbol{\varphi}_{N_2}(\varsigma_{{N_2}^{'}})\right]\in \mathbb{C}^{N \times N_1' N_2'}$,
and $\boldsymbol{\lambda}_k\in \mathbb{C}^{N_1' N_2'}$ represents the channel coefficients of $\mathbf{h}_k$ in the angular domain. Since the number of paths is usually limited, $\boldsymbol{\lambda}_k $ is essentially sparse, with each nonzero value being the corresponding composite path gain $\sqrt{\mu_k}\epsilon_i^k$. The exploitation of such a sparse structure has been shown to bring improved channel estimation performance and reduced complexity \cite{shaommv, power,lee,shaodim}.

\subsection{Signal Models}
\begin{figure}[t]
  \centering
\includegraphics [width=85mm] {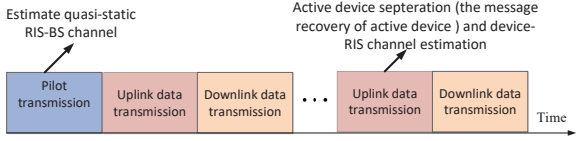}
\caption{The proposed joint active device separation and channel estimation frame structure for RIS-aided URA.}
\label{frame}
\end{figure}

According to the characteristics of RIS-aided URA, we propose a joint active device separation and channel estimation frame structure, as shown in Fig. \ref{frame}. Specifically, the BS first estimates the RIS-BS channel and then recovers the transmitted messages of the active devices and estimates the corresponding device-RIS channels. Finally, the BS conducts downlink data transmission based on the estimated CSI.
Since the BS and the RIS are placed at fixed positions,
the BS-RIS channel $\mathbf{U}$ is quasi-static \cite{beixiong}. As such, we only need to
estimate $\mathbf{U}$ once over a long period of time. The quasi-static RIS-BS channel can be efficiently estimated by the dual-link pilot transmission scheme proposed in \cite{sta}, where the BS transmits pilots to the RIS via the downlink channel with a single-antenna. Then, the RIS reflects the transmitted pilots back to the BS via the uplink channel with a set of predesigned reflection coefficients.
Based on the received pilots, the BS can accurately obtain the estimate of the quasi-static RIS-BS channel, i.e., $\hat{\mathbf{U}}$.

\begin{figure}[t]
  \centering
\includegraphics [width=85mm] {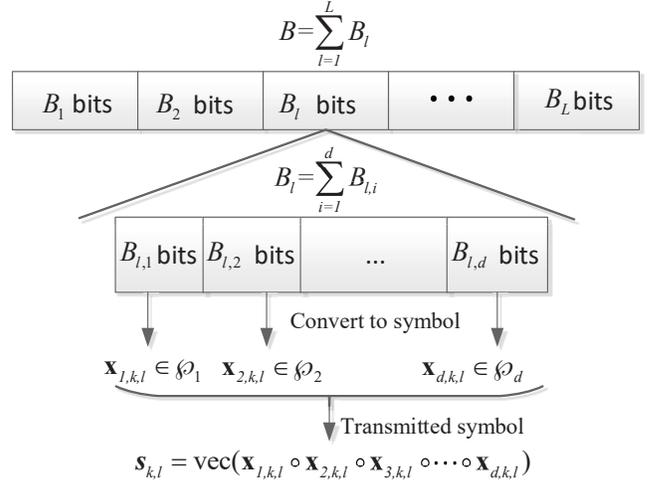}
\caption{Bit mapping of transmitted symbols for the $k$th device.}
\label{bit}
\end{figure}

On the other hand, a block transmission framework is adopted for joint active device separation and device-RIS channel estimation. Specifically, each active device divides its $B$-bit message
into $L$ sub-messages \cite{unsourced} and the length of the $l$-th sub-message is $B_l$ bits, as illustrated in Fig. \ref{bit}. Then, active devices transmit their $L$ sub-messages sequently, which are jointly recovered at the BS. Noting that since each subblock $l$ contains $K_a$ recovered sub-messages, only an instance of the transmitted data can be obtained in each subblock. However, the ultimate goal of the BS is to recover the whole set of $B$-bit messages that were transmitted by all the active devices. To achieve this goal, a decoder for decoding binary messages among different subblocks of all the $K_a$ active devices is needed. For example, in \cite{unsourced}, an outer tree-based decoder was adopted to stitch the decoded binary sequences across different subblocks (cf. Fig. \ref{URA}). Based on the recovered message, the BS can separate the active devices if they embed their identity (ID) information into the transmitted messages. Note that the inner decoder of URA is quite different from the conventional blind detection problem \cite{blind1}. First, the devices in URA share a common codebook, while the devices in blind detection may be assigned separate codebooks. Moreover, the receiver of URA neither knows the transmitted codewords, nor the number of transmitted codewords. In contrast, the blind detection generally has the knowledge of the number of transmitted codewords \cite{blind1}.

Following the URA paradigm \cite{unsourced}, we assume that all the devices in arbitrary subblock $l$ exploit the same constellation  $\wp=\{\mathbf{c}_1,\mathbf{c}_2,\cdots,\mathbf{c}_{2^{B_l}}\}$ containing $2^{B_l}$ elements. Considering the angular space channel in \eqref{angu3} and the RIS to the BS reflecting channel $\hat{\mathbf{U}}$, the received signals of the $l$th subblock at the BS can be expressed as
\begin{align}
\label{eqinter5}
\mathbf{Y}_l=\sum_{k=1}^{K_a}{\hat{\mathbf{U}}}\text{diag}(\mathbf{A}_\mathrm{R}\boldsymbol{\lambda}_k)\mathbf{v}_l\mathbf{s}_{k,l}^H+\mathbf{W}_l,
\end{align}
where $\mathbf{s}_{k,l}\in\wp \subset\mathbb{C}^{\tau}$ is the transmitted sequence of complex valued baseband symbols from the $k$th device over $\tau$
channel uses. $\mathbf{W}_l \in \mathbb{C}^{M\times \tau}$ is the additive white Gaussian noise. Herein,
the coefficient $[\mathbf{v}_l]_n=e^{\jmath \theta_{n}}, n\in\{1, \cdots, N\}$ denotes the phase shift of RIS reflecting element $n$ \cite{xu}. $\theta_{n}\in [0, 2\pi]$ and phase shift vector $\mathbf{v}_l$ is constant during the $l$-th subblock and varies from one subblock to another subblock.
Let $\mathbf{y}_l=\mathrm{vec}(\mathbf{Y}_l)$ and $\mathbf{w}_l=\mathrm{vec}(\mathbf{W}_l)$ denote the vectorized versions of $\mathbf{Y}_l\in \mathbb{C}^{M\times\tau}$ and $\mathbf{W}_l$, respectively. Moreover, we rearrange  $\hat{\mathbf{U}}\text{diag}(\mathbf{A}_{\mathrm{R}}\boldsymbol{\lambda}_k)\mathbf{v}_l$ as
$
\hat{\mathbf{U}}\text{diag}(\mathbf{A}_{\mathrm{R}}\boldsymbol{\lambda}_k)\mathbf{v}_l\!\!=\!\!\sum_{n=1}^{N}\!\hat{\mathbf{U}}(:,n)\{\mathbf{A}_{\mathrm{R}}\boldsymbol{\lambda}_k\} (n)\mathbf{v}_l(n)\!\!=\!\!\hat{\mathbf{V}}_l\mathbf{A}_{\mathrm{R}}\boldsymbol{\lambda}_k,
$
where $\hat{\mathbf{V}}_l=[\mathbf{v}_l(1)\hat{\mathbf{U}}(:,1),\mathbf{v}_l(2)\hat{\mathbf{U}}(:,2),\cdots,\mathbf{v}_l(N)\hat{\mathbf{U}}(:,N)]\in \mathbb{C}^{M\times N}$. Hence, \eqref{eqinter5} can be equivalently rewritten as
\begin{align}
\label{s5}
\mathbf{y}_l=\sum_{k=1}^{K_a}\mathbf{s}_{k,l}\otimes\mathbf{P}_l\boldsymbol{\lambda}_k+\mathbf{w}_l,
\end{align}
with $\mathbf{P}_l=\hat{\mathbf{V}}_l\mathbf{A}_\mathrm{R}\in \mathbb{C}^{M\times N}$. Based on the received signal $\mathbf{y}_l, l\in \{1,...,L\}$, the BS jointly estimates the channel coefficients $\boldsymbol{\lambda}_k$ and recovers the messages $\mathbf{s}_{k,l}$ for active device separation.

\section{Design of The RIS-Aided URA Architecture}
In this section, we first design the RIS passive reflection for facilitating RIS-assisted active device separation and channel estimation. Then, we formulate the joint active device separation and channel estimation problem as a high-order coupled tensor decomposition problem.
\subsection{RIS Reflection Design}
The phase-shift vector $\mathbf{v}_l, l=1,\cdots, L$, in \eqref{s5} controls the reflection of the waves impinging on the RIS. Given an optimized phase-shift design, an RIS can significantly improve the performance of the URA system. In the proposed RIS-aided URA system, the active device separation and the channel estimation are performed concurrently, thus we propose a specific optimized/random combined phase-shift design mechanism for $L$ subblocks. At the $l=1, \cdots, L-1$ stage, the phase shift is randomly generated, which can be regarded as a training process. At the last subblock, i.e., $l=L$, we design the optimal phase shift to enhance device separation and channel estimation.

Since the locations of the devices are a priori unknown, the ideal phase-shift design should provide a low misdetection probability for the entire coverage area. The authors in \cite{co} has verified that the probability of misdetection is a monotonically decreasing function with respect to the power of cascaded channel.

Under the multipath scenario \cite{emlihard}, the resultant IRS reflection
optimization problem is non-convex, and obtaining the global optimal
solution is in general intractable. As such, we adopt a compromised approach.
Considering practical delay-power spectral density in sparse channels, the first arrival path usually contains the most power of the channel due to short propagation distance \cite{power}. In other words, the channel gain is mainly determined by the first path. Motivated by such an observation, we formulate the worse-case optimization problem for maximizing the minimum channel gain over the first path among all devices, which is given by \cite{co}
\begin{align}
\label{OP3}
&\mathop{\max}\limits_{\{\mathbf{v}_L\}}\mathop{\min}\limits_{ k \in [1,K]} \frac{\varrho^2}{16\pi^2d_{\mathrm{DR},k}^2d_{\mathrm{RB}}^2}\left|\mathbf{a}^H_\mathrm{R}(\phi_1^k,\psi_1^{k})\text{diag}(\mathbf{a}_\mathrm{R}(\bar{\phi}_1,\bar{\psi}_1))\mathbf{v}_L\right|^2\nonumber\\
&\mathrm{s.t.}~~~[\mathbf{v}_L]_n=1,
n=1,\cdots,N,
\end{align}
where $d_{\mathrm{DR},k}$ denotes the distance from device $k$ to RIS and $d_{\mathrm{RB}}$ denotes the distance from the RIS to BS.

By setting $\bar{\mathbf{a}}^H(\phi_1^k,\psi_1^{k})=\mathbf{a}^H_\mathrm{R}(\phi_1^k,\psi_1^{k})\text{diag}(\mathbf{a}_\mathrm{R}(\bar{\phi}_1,\bar{\psi}_1))$
and $\bar{\mathbf{A}}(\phi_1^k,\psi_1^{k})=\bar{\mathbf{a}}(\phi_1^k,\psi_1^{k})\bar{\mathbf{a}}^H(\phi_1^k,\psi_1^{k})$, problem \eqref{OP3} can be reformulated as
\begin{align}
\label{OP33}
&\mathop{\max}\limits_{\{\mathbf{V}_L\}}\mathop{\min}\limits_{ k \in [1,K]} \left\{ \frac{\varrho^2}{16\pi^2d_{\mathrm{DR},k}^2d_{\mathrm{RB}}^2} \mathrm{Tr}\left(\bar{\mathbf{A}}(\phi_1^k,\psi_1^{k})\mathbf{V}_L\right)\right\}
\nonumber\\
&\mathrm{s.t.}~~~\mathbf{V}_L(n,n)=1,
n=1,\cdots,N,
\end{align}
where $\mathbf{V}_L=\mathbf{v}_L\mathbf{v}_L^H$. Note that the rank-one constraint on $\mathbf{V}_L$ is dropped in problem \eqref{OP33}. Fortunately, problem \eqref{OP33} is a standard convex semidefinite program, which can be solved by some optimization tools, i.e., CVX \cite{convex}.
Yet, $\mathrm{rank}(\mathbf{V}_L^{\mathrm{opt}})$ cannot be always guaranteed, where $\mathbf{V}_L^{\mathrm{opt}}$ is the optimal solution to problem \eqref{OP33}. In this context, it is difficult to obtain the optimal phase-shift vector
$\mathbf{v}_L^{\mathrm{opt}}$ from $\mathbf{V}_L^{\mathrm{opt}}$. To solve this problem, the following Gaussian randomization approach is applied to $\mathbf{V}_L^{\mathrm{opt}}$ \cite{mosek}:
\begin{enumerate}
\item For $j \in \{0, 1, \cdots, J-1\}=\mathcal{J}$, generate $J$ random vectors $\boldsymbol{\nu}_j\sim \mathcal{CN}_{N}(\boldsymbol{\nu}_j|\mathbf{0},\mathbf{V}_L^{\mathrm{opt}})$
    and set $|[\boldsymbol{\nu}_j]_n|=1,\forall n, \forall j$.

\item Search $\hat{\mathbf{v}}_L^{\mathrm{opt}}=\arg\mathop{\max}\limits_{\boldsymbol{\nu}_j ,\forall j\in \mathcal{J}}\mathop{\min}\limits_{ k \in \mathcal{Q}}\left\{\boldsymbol{\nu}_j^H\bar{\mathbf{A}}(\phi_1^k,\psi_1^{k})\boldsymbol{\nu}_j\right\}$.
\end{enumerate}
Thus, we can obtain a sub-optimal phase-shift vector $\hat{\mathbf{v}}_L^{\mathrm{opt}}$.

\subsection{Coupled Tensor Modeling}
Based on the designed phase shift $\hat{\mathbf{v}}_l, l=1,\cdots,L$, the data symbol is transmitted from the active devices. In this paper, we construct the constellation structure $\wp$ according to the tensor decomposition format \cite{newT}. As shown in Fig. \ref{bit}, it is assumed that the channel use, i.e., $\tau$, can be factorized as $\tau=\Pi_{i}^{d}\tau_i$ for some $d\geq2$ and $\tau_i\geq 2, \forall i $. Subsequently, the complex symbols transmitted by the $k$th device at the $l$th subblock, denoted by $\mathbf{s}_{k,l} \in \mathbb{C}^{\tau}$, can be rewritten as the vector representation of a rank-$1$ tensor $\mathcal{S}_{k,l} \in \mathbb{C}^{\tau_1\times\tau_2\times\cdots\times \tau_d}$ of dimensions $\tau_1,\tau_2,\cdots,\tau_d$, that is
\begin{align}\label{cons}
\mathbf{s}_{k,l}=\mathrm{vec}(\mathcal{S}_{k,l})\in \mathbb{C}^{\Pi_{i}^{d}\tau_i}=\mathbb{C}^{\tau},
\end{align}
with
$
\mathcal{S}_{k,l}=\mathbf{x}_{1,k,l}\circ \mathbf{x}_{2,k,l}\circ \cdots \circ \mathbf{x}_{d,k,l}, \forall k,l$,
where each $\mathbf{x}_{i,k,l}$ is generated from a sub-constellation $\wp_i$, which is defined as a discrete subset of $\mathbb{C}^{\tau_i}$. As shown in Fig. \ref{bit}, bit information can be mapped to symbol as follows. $B_l$ coded bits are split into $d$ sets of $\{B_{1,1}\},...,\{B_{l,d}\}$ bits, respectively, corresponding to the $d$ tensor dimensions. For the $i$th set, $B_{l,i}$-bit data is mapped to an element $\mathbf{x}_{i,k,l}$. Then, the vector symbol $\mathbf{s}_{k,l}$ is formed by the calculation in \eqref{cons}. Actually, all possible combinations of elements in the sub-constellations result in the vector constellation $\wp \subset\mathbb{C}^{\tau}$, which is given by $\wp=\left \{\mathrm{vec}(\mathbf{x}_{1}\circ \mathbf{x}_{2}\circ \cdots \circ \mathbf{x}_{d}), \mathbf{x}_{1}\!\in\!\wp_1,\!\cdots,\! \mathbf{x}_{d}\in\wp_d  \right \}$.
\begin{figure}[t]
  \centering
  \setlength{\abovecaptionskip}{0.cm}
\includegraphics [width=0.48\textwidth]{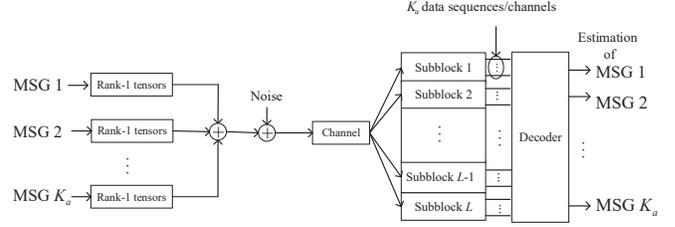}
\caption{High-level description of the proposed RIS-based unsourced random access scheme, where MSG $k$ denotes the message of the $k$th device.}
\label{URA}
\end{figure}
Substituing \eqref{cons} into \eqref{s5}, we can reformulate \eqref{s5} as the following equivalent tensor decomposition form
\begin{align}\label{e1}
\!\!\!\!\!\mathcal{Y}_l\!=\!\underbrace{\sum_{k=1}^{K_a}\mathbf{x}_{1,k,l}\circ \mathbf{x}_{2,k,l}\circ \mathbf{x}_{3,k,l}\circ \cdots \circ \mathbf{x}_{d,k,l}\circ \mathbf{P}_l\boldsymbol{\lambda}_k}_{\mathcal{Y}_l^0}+\mathcal{W}_l,
\end{align}
where $\mathcal{W}_l \in \mathbb{C}^{\tau_1\times\cdots\tau_d\times M}$ is the additive white Gaussian noise represented in the tensor space. Therefore, in \eqref{e1}, each active device transmits a sequence that is associated with a rank-one tensor of order $d$, while $\mathcal{Y}_l^0$ is the noise-free tensor of order
$d +1$ and has dimensions $\tau_1, \cdots ,\tau_d ,M$, respectively. Its tensor rank is at most $K_a$, since a tensor is said to be rank-$K_a$ whenever $K_a$ is the smallest integer such that the tensor can be written as a sum of $K_a$ rank-1 tensors \cite{tenrank}.

Consequently, we can formulate the design of joint active device separation and channel estimation as the following non-convex optimization problem:
\begin{align}\label{eE2}
&\mathop {\arg\min}\limits_{\mathbf{x}_{i,k,l}\in \wp_i^{\tau_i},\boldsymbol{\lambda}_k\in\mathbb{C}^N}\sum_{l=1}^{L}\left\| \mathcal{Y}_l-\sum_{k=1}^{K_a}\mathbf{x}_{1,k,l}\circ \cdots \circ \mathbf{x}_{d,k,l}\circ (\mathbf{P}_l\boldsymbol{\lambda}_k) \right \|_F^2\nonumber\\
&\mathrm{s. t.}~\|\boldsymbol{\lambda}_k\|_{0}\leq \zeta_s,~~k=1,2,\cdots,K_a,
\end{align}
where $\zeta_s$ is a predefined parameter for imposing the channel sparsity. By solving problem \eqref{eE2}, we can obtain $\mathbf{x}_{1,k,1},\cdots,\mathbf{x}_{d,k,L}$ and $\boldsymbol{\lambda}_k$ for $k=1,2,\cdots,K$. After that, the estimation of transmitted data symbol $\hat{\mathbf{s}}_{k,l}$ can be formed by computing the vector outer product according to \eqref{cons}. The high-level description of the proposed RIS-based URA scheme is shown in Fig. \ref{URA}.

\subsection{The Challenges Ahead}
For the proposed RIS-aided URA architecture, the key is to solve problem \eqref{eE2}. However, there are two major challenges that prohibit the straightforward application of some existing tensor decomposition advances (e.g., using either optimization methods in \cite{cou} and \cite{cou2} or applying the Bayesian framework in \cite{chengpro} and \cite{cheng1}):
\begin{enumerate}
\item Problem \eqref{eE2} is a non-standard coupled tensor decomposition problem as the sparse profile vector is common and nonlinearly coupled with other parameters in modeling multiple tensors $\mathcal{Y}_l, \forall l$. In contrast, in \cite{cou} and \cite{cou2}, the coupled factor matrix $\mathbf{B}_k$ is common for $\forall l \in \{1,\cdots,L\}$ and does not pose a sparse structure. However, in problem \eqref{eE2}, the coupled information is a sparse vector $\boldsymbol{\lambda}_k$ come from $\mathbf{B}_k=\mathbf{P}_l\boldsymbol{\lambda}_k$ with known $\mathbf{P}_l$. The exploitation of such a sparse structure can be exploited to improve channel estimation performance and to reduce the algorithm complexity.

\item In \cite{cou} and \cite{cou2}, the authors assumed that the tensor rank $K_a$ is known. However, the tensor rank (number of active devices $K_a$) in problem \eqref{eE2} is random and unknown in practice introducing NP-hardness to the estimation based on tensor data. The accurate information about the model complexity is essential in avoiding overfitting of noises or underfitting of signals. Therefore, we introduce the variational inference algorithm that can automatically learn both the transmit data, the channel state information, and the active device number from the measured data at the BS under the Bayesian learning framework.
\end{enumerate}

\section{Coupled Tensor-Based Automatic Detection Algorithm: Bayesian Learning}
Note that solving the discrete problem in \eqref{eE2} optimally via an exhaustive search, however, requires $2^{K_aB}$ number of evaluations of the objective function. To circumvent the complexity issue, we first relax the discrete domain ${{\{\mathbf{x}_{i,k,l}\in \wp_i^{\tau_i}\}_{i=1}^{d}}_{k=1}^{K_a}}_{l=1}^{L}$ in \eqref{eE2} to a continuous one ${{\{\mathbf{x}_{i,k,l}\in \mathbb{C}^{\tau_i}\}_{i=1}^{d}}_{k=1}^{K_a}}_{l=1}^{L}$. Besides, $K_a$ is unknown for the modeling of all $\mathcal{Y}_l$, $l=1,2,\cdots,L$, and its estimation has been shown to be NP-hard. To tackle this challenge, an effective scheme is to introduce two regularization terms \cite{zhao} that penalize the model
complexity and avoid possible overfitting of noise as follows
\begin{align}\label{eT}
&\mathop {\arg\min}\limits_{{\{\mathbf{X}_l^i\in\mathbb{C}^{\tau_i \times K}\}_{i=1}^{d}}_{l=1}^{L}\mathbf{G}\in\mathbb{C}^{M\times K}}\sum_{l=1}^{L}\left\| \mathcal{Y}_l-\left[\!\!\left[\mathbf{X}_l^1,\cdots,\mathbf{X}_l^d, ~\mathbf{P}_l\mathbf{G}\right]\!\!\right]\right \|_F^2\nonumber\\
&\!+\!\mathop \sum \limits_{k=1}^{K} \gamma_k\!\left(\!\sum_{l=1}^{L}\mathop \sum \limits_{i=1}^{d}\mathbf{X}_l^{i}(:,k)^H\mathbf{X}_l^i(:,k)\!\right)\!+\!\mathop \sum \limits_{k=1}^{K} \eta_k\mathbf{G}(:,k)^H\mathbf{G}(:,k)\nonumber\\
&\mathrm{s. t.}~\|\mathbf{G}(:,k)\|_{0}\leq \zeta_s,~~k=1,2,\cdots,K,
\end{align}
where $\mathbf{X}_l^i \in\mathbb{C}^{\tau_i \times K}$ with the $k$th column being $\mathbf{x}_{d,k,l}$, $\mathbf{G} \in\mathbb{C}^{M \times K}$ is defined similarly but with the $k$th column being $\boldsymbol{\lambda}_k$.
Herein, we set the column number of all factor matrices $\{\mathbf{X}_l^1, \cdots, \mathbf{X}_l^d, \mathbf{G}\}$ as $K$, which is the maximum possible number of active devices $K_a$.
The idea about regularization term is that if $\gamma_k>0$ and $\eta_k>0$ are sufficiently large (e.g., $10^3$) after inference, the elements of the $k$th columns in the optimal ${\{\mathbf{X}_l^{i}\}_{l=1}^L}_{i=1}^d$ and $\mathbf{G}$ approach zeros. Then, the corresponding column can be pruned out and the number of remaining columns with significant non-zero values in each factor matrix returns an estimate of number of active devices.
However, the choice of regularization parameters $\gamma_k$ and $\eta_k$ is not straightforward and computationally demanding, since setting $\gamma_k$ and $\eta_k$ too large would lead to excessive residual squared error, while setting them too small may incur overfitting of noises. Therefore, we develop an intelligent algorithm that can automatically learn both the factor matrices and regularization parameters from the measured data at the BS under the Bayesian learning framework.

\subsection{Element-Wise Sparsity and Low-Rank Inducing Probabilistic Modeling}
Firstly, to apply the Bayesian learning framework, a probabilistic model, which encodes the knowledge of problem \eqref{eT}, needs to be established. To strike the trade-off between the flexibility of knowledge encoding and the tractability of the inference algorithm, we propose a novel probabilistic model by interpreting each term in problem \eqref{eT} via some probability density functions (pdfs) \cite{zhao}. We start with the last regularization term in the objective function of \eqref{eT}, which can be modeled as a circularly-symmetric complex Gaussian prior distribution\footnote{This is inspired by the $l_2$ norm in the last regularization term and the assumption that device to RIS channel is independent among the different devices.} of the columns in matrix $\mathbf{G}$, i.e., $\prod_{k=1}^{K}\mathcal{CN}
\left(\mathbf{G}(:,k)|\mathbf{0},(\eta_k)^{-1}\mathbf{I}\right)$. On the other hand, note that the $l_0$-norm constraint in \eqref{eT} is imposed such that the elements in each column of channel are also sparse. This observation inspires a sparsity-aware probabilistic modeling on each element of $\mathbf{G}$. Therefore, by taking both the column-wise sparsity (i.e., low-rank) and the element-wise sparsity structure into account, we propose the following novel prior for matrix $\mathbf{G}$:
\begin{align}\label{pen3}
&p\left(\mathbf{G}|\{\eta_k\}_{k=1}^{K},{\{\boldsymbol{\xi}(n,k)\}_{n=1}^{N}}_{k=1}^{K}\right)\nonumber\\
&=\prod_{k=1}^{K}\!\left[\mathcal{CN}
\left(\mathbf{G}(:,k)|\mathbf{0},\eta_k^{-1}\mathbf{I}\right)\right.\left.\prod_{n=1}^{N}\mathcal{CN}\left(\!\mathbf{G}(n,k)|0,\boldsymbol{\xi}(n,k)^{-1}\right)\right],
\end{align}
with
\begin{align}\label{etap}
p(\{\eta_k\}_{k=1}^{K}|\boldsymbol{\imath}_{\eta})&=\prod_{k=1}^{K}\!\!\mathrm{gamma}(\eta_k|\delta,\delta)\!\!=\!\!\prod_{k=1}^{K}\!\!\eta_k^{\delta-1}\exp(-\delta\eta_k),
\end{align}
\begin{align}\label{etapo}
&p({\{\boldsymbol{\xi}(n,k)\}_{n=1}^{N}}_{k=1}^{K}|\boldsymbol{\imath}_{\boldsymbol{\xi}})=\prod_{n=1}^{N}\prod_{k=1}^{K}\mathrm{gamma}(\boldsymbol{\xi}(n,k)|\delta,\delta)\nonumber\\
&=\prod_{n=1}^{N}\prod_{k=1}^{K}\boldsymbol{\xi}(n,k)^{\delta-1}\exp(-\delta\boldsymbol{\xi}(n,k)),
\end{align}
where $\delta>0$ is a small number that indicates the non-informativeness of the prior model, the natural parameter $\boldsymbol{\imath}_{\eta}=[-\delta\mathbf{1}_{K};(\delta-1)\mathbf{1}_{K}]$, and $\boldsymbol{\imath}_{\boldsymbol{\xi}}=[-\delta\mathbf{1}_{NK};(\delta-1)\mathbf{1}_{NK}]$.

The proposed prior in \eqref{pen3} is employed to model the sparsity pattern of $\mathbf{G}$.
Different from the previous works \cite{chengpro, zhao} that only adopted Gaussian-gamma pairs for modeling the column-wise sparsity, the prior in \eqref{pen3} utilizes the product rule of probability to further encode the element-wise sparsity information. The proposed prior has a clear physical interpretation. Particularly, $\boldsymbol{\xi}(n,k)^{-1}$ can be interpreted as the power of each element $\mathbf{G}(n,k)$,
while $\eta_k^{-1}$ has a physical interpretation as the power of each column in $\mathbf{G}$.
Therefore, if $\eta_k^{-1}$ is learnt to approach zero, regardless of $\boldsymbol{\xi}(n,k)$, the corresponding columns in the factor matrices play no role in channel modeling and thus can be pruned out by thresholding. Similarly, for a nonzero column ($\eta_{k}^{-1}\neq 0$), $\mathbf{G}(n,k)$ would be shrunk to zero as $\boldsymbol{\xi}(n,k)^{-1}$ goes to zero. Therefore, the proposed prior in \eqref{pen3} can simultaneously promote element-wise sparsity of the channel and low-rank property of tensor data, which mimic the sparsity constraint and the last regularization term in \eqref{eT}.
Note that the conventional way to promote both element-wise and column-wise sparsity is to use hierarchical Bayesian \cite{vi}. However, the resulting penalty is less capable in enforcing a structural low-rank to the tensor. In this context, multiplying two priors serves as a viable approach since both hyperparameters influence $\mathbf{G}$ directly.

Similarly, for the regularization term about $\mathbf{X}_l^i$ in problem \eqref{eT}, it can be modeled as a zero-mean circularly-symmetric complex Gaussian prior distribution over the columns of the factor matrices as follows \cite{chengpro}.
\begin{align}\label{peX}
p\left({\{\mathbf{X}_l^i\}_{l=1}^{L}}_{i=1}^{d}|\{\gamma_k\}_{k=1}^{K}\right)=\prod_{l=1}^{L}\prod_{i=1}^{d}\prod_{k=1}^{K}\mathcal{CN}
\left(\mathbf{X}_{l}^i(:,k)|\mathbf{0},\gamma_k^{-1}\mathbf{I}\right).
\end{align}
A gamma distribution is adopted for the penalty parameter $\gamma_k$ to enforce the column-wise sparsity \cite{chengpro}:
\begin{align}\label{gammap}
\!\!\!\!\!p(\{\gamma_k\}_{k=1}^{K}|\boldsymbol{\imath}_{\gamma})\!=\!\prod_{k=1}^{K}\!\!\mathrm{gamma}(\gamma_k|\delta,\delta)\!\!=\!\!\prod_{k=1}^{K}\!\!\gamma_k^{\delta-1}\exp(-\delta\gamma_k),
\end{align}
where $\gamma_k^{-1}$ can be interpreted as the power of each column in $\mathbf{X}_l^i$ and $\boldsymbol{\imath}_{\gamma}=[-\delta\mathbf{1}_{K};(\delta-1)\mathbf{1}_{K}]$.
Noting that after integrating the gamma hyper-prior, the marginal distribution of model parameters \eqref{peX} is a Student's $t$ distribution, which is strongly peaked at zero and with heavy tails, thus promoting sparsity. Furthermore, the gamma hyper-prior \eqref{gammap} is conjugate to the Gaussian prior \eqref{peX}. This conjugacy permits the closed-form solution of the variational inference \cite{vi}.

Finally, since the elements of the additive noise $\mathcal{W}_l$ obeys white Gaussian distribution, the sum of the squared error term in problem \eqref{eT} can be interpreted as the negative log of a likelihood function given by
\begin{align}\label{y}
&p(\{\mathcal{Y}_l\}_{l=1}^{L}|{\{\mathbf{X}_l^i\}_{i=1}^{d}}_{l=1}^{L}, \mathbf{G}, \beta)\nonumber\\
&\propto \exp\left(-\beta \mathop \sum \limits_{l=1}^{L}\left\| {\mathcal{Y}_l}\right.\right.\left.\left.-\left[\!\!\left[\mathbf{X}_l^1,~\mathbf{X}_l^2,\cdots,\mathbf{X}_l^d, ~\mathbf{P}_l\mathbf{G}\right]\!\!\right] \right\|_F^2\right),
\end{align}
where the noise precision $\beta$ obeys gamma distribution, i.e., $p(\beta|\boldsymbol{\imath}_{\beta})\propto\beta^{\delta-1}\exp(-\delta\beta)$ with natural parameters $\boldsymbol{\imath}_{\beta}=[-\delta;(\delta-1)]$.
\begin{figure*}[t]
  \centering
\includegraphics [width=0.8\textwidth]{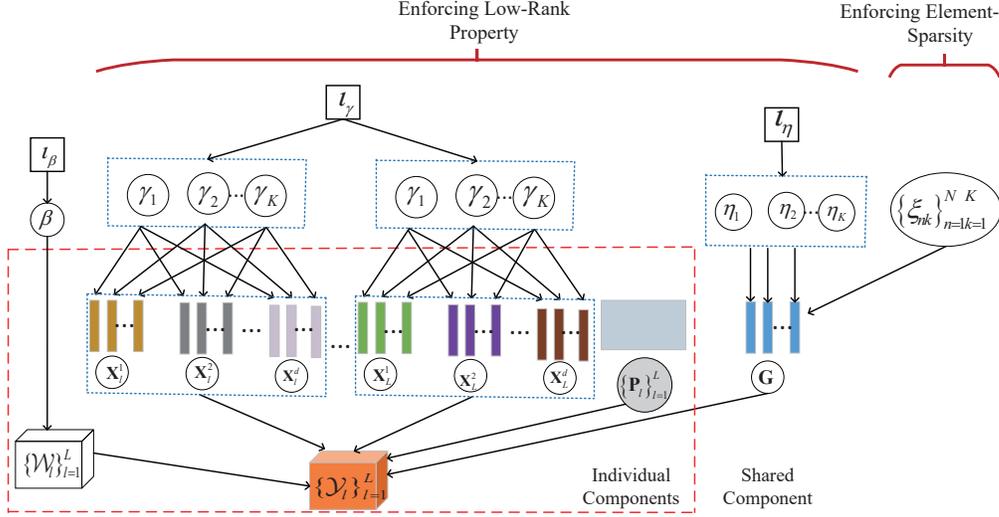}
\caption{The proposed probabilistic model for problem \eqref{eT} that incorporates the information of both element-wise sparsity and low-rank property. Filled-in circle and cube indicate the observable variables, hollow circles and cube denote the unknown variables, the boxes denote pre-specified hyperparameters, and arrows describe conditional dependencies between variables.}
\label{babilistic3}
\end{figure*}

Graphical illustration of the proposed complete probabilistic model is presented in Fig. \ref{babilistic3}. Let $\boldsymbol{\Theta} =\{{{\{\mathbf{X}_l^i\}_{i=1}^{d}}_{l=1}^{L}}, \mathbf{G}, \beta,{\{\gamma_k\}_{k=1}^{K}},{\{\eta_k\}_{k=1}^{K}},{\{\boldsymbol{\xi}(n,k)\}_{n=1}^{N}}_{k=1}^{K}\}$ collect all the unknown random variables, the joint pdf of $\boldsymbol{\Theta}$ and $\{\mathcal{Y}_l\}_{l=1}^L$ is given as follows:
\begin{figure*}
\begin{align}\label{joint}
\!\!&p(\boldsymbol{\Theta} ,\{\mathcal{Y}_l\}_{l=1}^L)=p(\{\mathcal{Y}_l\}_{l=1}^{L}|{{\{\mathbf{X}_l^i\}_{i=1}^{d}}_{l=1}^{L}}, \mathbf{G}, \beta) p({{\{\mathbf{X}_l^i\}_{i=1}^{d}}_{l=1}^{L}}|{\{\gamma_k\}_{k=1}^{K}})p\!\left(\!\mathbf{G}|\{\eta_k\}_{k=1}^{K},{\{\boldsymbol{\xi}(n,k)\}_{n=1}^{N}}_{k=1}^{K}\!\right)\!
\nonumber\\
\!\!&\times p\!\left(\!\beta|\boldsymbol{\imath}_{\beta}\!\right)\!p\!\left(\!{\{\gamma_k\}_{k=1}^{K}}|\boldsymbol{\imath}_{\gamma}\!\right)\!p\!\left(\!{\{\eta_k\}_{k=1}^{K}}|\boldsymbol{\imath}_{\eta}\!\right)\!p\!\left(\!{\{\boldsymbol{\xi}(n,k)\}_{n=1}^{N}}_{k=1}^{K}|\boldsymbol{\imath}_{\boldsymbol{\xi}}\!\right)\!\nonumber\\
\!\!&\propto  \exp\left\{-\beta\sum_{l=1}^{L}\left \| \mathcal{Y}_l-\left[\!\!\left[\mathbf{X}_l^1,~\mathbf{X}_l^2,\cdots,\mathbf{X}_l^d, ~\mathbf{P}_l\mathbf{G}\right]\!\!\right]\right \|_F^2\right.\left.\!-\!\sum_{l=1}^{L}\sum_{i=1}^{d}\mathrm{Tr}(\boldsymbol{\Lambda}\mathbf{X}_{l}^{i,H}\mathbf{X}_{l}^i)\!+\!L\left(\Pi_{i=1}^{d+1}\tau_iM\right)\ln\beta\right.\nonumber\\
\!\!&\left.\!-\!\delta\beta\!-\!\sum_{k=1}^{K}\delta\gamma_k+(\delta\!-\!1)\ln\beta-\mathrm{Tr}(\bar{\boldsymbol{\Lambda}}\mathbf{G}^H\mathbf{G})\!-\!
\sum_{k=1}^{K}\mathrm{Tr}(\mathbf{Z}_k\mathbf{G}(:,k)\mathbf{G}(:,k)^H)\!+\!(N\!+\!\delta\!-\!1)\sum_{k=1}^{K}\ln{\eta_k}\right.\nonumber\\
\!\!&\left.\!-\!\sum_{k=1}^{K}\delta\eta_k\!-\!\delta\sum_{n=1}^{N}\sum_{k=1}^{K}{\boldsymbol{\xi}(n,k)}\!+\!(N\!+\!\delta\right.\left.\left.\!-\!1)\sum_{n=1}^{N}\sum_{k=1}^{K}\ln{\boldsymbol{\xi}(n,k)}\!+\!\left(\!L\sum_{i=1}^{d}\tau_i+\delta\!-\!1\!\right)\sum_{k=1}^{K}\ln{\gamma_k}\right.\!\!\right\}\!, \!\!\!
\end{align}
\hrulefill
\end{figure*}
with
$
\mathbf{Z}_k=\text{diag}(\boldsymbol{\xi}({1,k}),\boldsymbol{\xi}({2,l}),\cdots,\boldsymbol{\xi}(N,k)\in \mathbb{C}^{N\times N}$,
$\boldsymbol{\Lambda}\!=\!\text{diag}(\gamma_1,\!\cdots,\!\gamma_K\!)\in \mathbb{C}^{K\times K}$, and $\bar{\boldsymbol{\Lambda}}\!=\!\text{diag}(\eta_1,\!\cdots,\!\eta_K\!)\in \mathbb{C}^{K\times K}.
$
Given the probabilistic model $p(\boldsymbol{\Theta} ,\{\mathcal{Y}_l\}_{l=1}^L)$ in \eqref{joint} at the top of next page, the next goal of Bayesian inference is to learn the model parameters $\boldsymbol{\Theta}$ from the tensor data $\{\mathcal{Y}_l\}_{l=1}^L$, where the posterior distribution of unknown model parameters, i.e., $p(\boldsymbol{\Theta}|\{\mathcal{Y}_l\}_{l=1}^L)$, is needed to be sought before further proceeding. Noting that maximizing the posterior probability $p(\boldsymbol{\Theta}|\{\mathcal{Y}_l\}_{l=1}^L)=p\left(\boldsymbol{\Theta},\{\mathcal{Y}_l\}_{l=1}^L\right)/p\left(\{\mathcal{Y}_l\}_{l=1}^L\right)$ is similar to addressing the problem \eqref{eT}. The difference is that the problem \eqref{eT} cannot learn the regularization parameters.
\subsection{Algorithm Development via Variational Inference}
Unfortunately, the probabilistic model described by the joint pdf \eqref{joint} is still sophisticated. In particular, solving the multiple integrals for computing the posterior distribution is generally intractable. To address this problem, we adopt the variational inference method which establishes a variational distribution $Q(\boldsymbol{\Theta})$ to approximate the true posterior $p(\boldsymbol{\Theta}|\{\mathcal{Y}_l\}_{l=1}^L)$. To achieve this goal, $Q(\boldsymbol{\Theta})$ is the solution which minimizes the Kullback-Leibler (KL) divergence, i.e.,
\begin{align}\label{kl}
&\mathop\mathrm{minimize} \limits_{Q(\boldsymbol{\Theta})}~\mathrm{KL}(Q(\boldsymbol{\Theta})|p(\boldsymbol{\Theta}|\{\mathcal{Y}_l\}_{l=1}^L))
\triangleq \mathop\mathrm{minimize} \limits_{Q(\boldsymbol{\Theta})}~ \nonumber\\
&-\mathbb{E}_{Q(\boldsymbol{\Theta})}\left\{\ln \frac{p(\boldsymbol{\Theta}|\{\mathcal{Y}_l\}_{l=1}^L)}{Q(\boldsymbol{\Theta})}\right\}.
\end{align}

To address the problem \eqref{kl}, mean-field approximation \cite{vi} is widely used to offer a tractable solution of \eqref{kl}, which assumes that the variational pdf can be represented in a fully factorized form, i.e., $Q(\boldsymbol{\Theta})= \mathop \prod \limits_{j=1}^{J}Q({\Theta}_j)$, where ${\Theta}_j\in \boldsymbol{\Theta}$ is part of $\boldsymbol{\Theta}$ with $\bigcup_{j=1}^{J}\boldsymbol{\Theta}_j=\boldsymbol{\Theta}$ and $\bigcap_{j=1}^{J}\boldsymbol{\Theta}_j=\varnothing$, and $J$ is the number of subsets. With this factorization, \eqref{kl} reduces to
\begin{align}\label{kl1}
\mathop\mathrm{minimize} \limits_{Q(\{{\Theta}_j\}_{j=1}^J)}-\mathbb{E}_{\{Q({\Theta}_j)\}_{j=1}^J}\left\{\ln \left(\frac{p(\boldsymbol{\Theta}|\{\mathcal{Y}_l\}_{l=1}^L)}{\prod_{j=1}^{J}Q({\Theta}_j)}\right)\right\}.
\end{align}

The factorized structure of $\{Q(\Theta_j)\}_{j=1}^J$ motivates the use of block coordinate descent to obtain a suboptimal solution of \eqref{kl1}. Particularly,
by fixing the variational pdfs $\{Q(\Theta_i)\}_{i\neq j}$, $Q(\Theta_j)$ is optimized as follows:
\begin{align}\label{kl2}
\mathop\mathrm{minimize} \limits_{Q({\Theta}_j)}~&\int Q({\Theta}_j)\left(-\mathbb{E}_{\prod_{i\neq j}Q({\Theta}_i)}\ln p(\boldsymbol{\Theta},\{\mathcal{Y}_l\}_{l=1}^L)\right.\nonumber\\
&\left.+\ln Q({\Theta}_j)\right)d{\Theta}_j.
\end{align}
It has been shown in \cite{vi} that the optimal solution of \eqref{kl2} is given by
\begin{align}\label{kl3}
Q^\dagger({\Theta}_j)=\frac{\exp(\mathbb{E}_{\prod_{i\neq j}Q({\Theta}_i)}\ln p(\boldsymbol{\Theta},\{\mathcal{Y}_l\}_{l=1}^L))}{\int \mathbb{E}_{\prod_{i\neq j}Q({\Theta}_i)}\ln p(\boldsymbol{\Theta},\{\mathcal{Y}_l\}_{l=1}^L)d\Theta_j}, \forall j.
\end{align}
Using \eqref{kl3}, we derive the closed-form posterior update for each variational pdf, i.e., $Q^\dagger({\Theta}_j)$ in the following subsections.

\subsection{Learning Device-RIS Channel: Inferring Coupled Factor $Q(\mathbf{G})$}
We start by deriving the explicit expression of the optimal variational pdf for the coupled factor $Q^\dagger(\mathbf{G})$, whose mean matrix gives the MMSE estimate of the interested device-RIS channel matrix \cite{ste}. It is clear that the likelihood function
in \eqref{y} results in complicated couplings among the shared factor matrix $\mathbf{G}$, which make the derivation of $Q^\dagger(\mathbf{G})$ challenging. To overcome this difficulty, we first define $\mathcal{Y}_l(d+1)\in\mathbb{C}^{M \times \tau_1\tau_2\cdots\tau_d}$ as the unfolding operation \cite{zhao} on a $(d+1)$th-order tensor
$\mathcal{Y}_l\in\mathbb{C}^{\tau_1\times\cdots\times\tau_d\times M}$ along its $(d+1)$-th mode. By substituting \eqref{joint} into \eqref{kl3}, using the result $\left\|\mathbf{A}\right\|_F^2=\mathrm{Tr}(\mathbf{A}\mathbf{A}^H)$ and then only keeping the terms relevant to $\mathbf{G}$, we obtain
\begin{figure*}
\begin{align}
&Q^\dagger(\mathbf{G})\propto \exp \left\{\mathbb{E}\left[-\beta\sum_{l=1}^{L}\left \| \mathcal{Y}_l-\left[\!\!\left[\mathbf{X}_l^1,\cdots,\mathbf{X}_l^d, ~\mathbf{P}_l\mathbf{G}\right]\!\!\right]\right \|_F^2-
\sum_{k=1}^{K}\mathrm{Tr}(\mathbf{Z}_k\mathbf{G}(:,k)\mathbf{G}(:,k)^H)\right.\right.\nonumber\\
&\left.\left.- \mathrm{Tr}(\bar{\boldsymbol{\Lambda}}\mathbf{G}^H\mathbf{G})\right] \right\}\propto \! \exp \! \left\{\right.\left.\!-\mathrm{Tr}(
\sum_{l=1}^{L}\mathbf{P}_l^H\mathbf{P}_l\mathbf{G}\underbrace{\mathbb{E}[\beta]\mathbb{E}\left[\left(\!\mathop \diamond \limits_{i=1}^{d}\mathbf{X}_{l}^i\!\right)^T\!\left(\!\mathop \diamond \limits_{i=1}^{d}\mathbf{X}_{l}^i\!\right)^*\right]}_{\boldsymbol{\Xi}_l^{-1}}\mathbf{G}^H\!-\!\mathbf{P}_l^H\mathbf{P}_l\right.\nonumber\\
&\left.\left.\times\underbrace{\mathbb{E}[\beta](\mathbf{P}_l^H\mathbf{P}_l)^{-1}\mathbf{P}_l^{H}\mathcal{Y}_l(d+1)\left(\mathop \diamond \limits_{i=1}^{d}\mathbb{E}[\mathbf{X}_{l}^i]\right)^*\boldsymbol{\Xi}_l}_{\boldsymbol{\Phi}_l }\boldsymbol{\Xi}_l ^{-1}\mathbf{G}^H-\mathbf{P}_l^H\mathbf{P}_l\mathbf{G}\boldsymbol{\Xi}_l ^{-1}{\boldsymbol{\Phi}_l^H })\right\}\right.\nonumber\\
&\times\exp \left\{\!-\!
\mathrm{vec}(\mathbf{G}^H)^H(\mathbb{E}[\boldsymbol{\Gamma}]\!+\!\mathbb{E}[\mathbf{Z}])\mathrm{vec}(\mathbf{G}^H)\!\right\}\label{opg},
\end{align}
\hrulefill
\end{figure*}
with
$
\mathbf{Z}\!\!=\!\!\text{diag}(\boldsymbol{\xi}(1,1),\!\cdots\!,\boldsymbol{\xi}(1,K),\!\cdots\!,\boldsymbol{\xi}(N,1),\!\cdots\!,\boldsymbol{\xi}(N,K))\!\in \! \mathbb{C}^{NK\times NK}
$,
and
$\boldsymbol{\Gamma}=\text{diag}(\mathbf{1}_{N}\otimes \left[\eta_1,\cdots,\eta_K\right])\in \mathbb{C}^{NK\times NK}$,
where $\mathop \diamond \limits_{i=1}^{d}\mathbf{X}_{l}^i=\mathbf{X}_{l}^d\mathop \diamond \limits\mathbf{X}_{l}^{d-1}\mathop \diamond \limits\cdots\mathop \diamond \limits\mathbf{X}_{l}^1$ denotes the multiple Khatri-Rao products. Let $\mathbf{A}=(\mathbf{a}_1,\cdots,\mathbf{a}_n) \in \mathbb{C}^{K\times N}$ with $\mathbf{a}_n\in \mathbb{C}^K$. The vectorization operation on $\mathbf{A}$ is expressed by $\mathrm{vec}(\mathbf{A})=(
\mathbf{a}_1^T,\cdots,\mathbf{a}_n^T)^T\in \mathbb{C}^{NK}$.

However, there are some complicated matrix multiplications in the last three lines of formula \eqref{opg} at the top of this page arising from $\mathbf{P}_l, \forall l$, which make it difficult to identify the variational pdf $Q^\dagger(\mathbf{G})$. To reveal its structure, we utilize the property of the vectorization operation that $\mathrm{Tr}(\mathbf{A}\mathbf{X}\mathbf{B}\mathbf{X}^T)=\mathrm{vec}(\mathbf{X})^T(\mathbf{B}\otimes\mathbf{A}^T)\mathrm{vec}(\mathbf{X})$ with $\mathbf{X}\in \mathbb{C}^{m\times n}$, $\mathbf{A}\in \mathbb{C}^{m\times m}$ and $\mathbf{B}\in \mathbb{C}^{n\times n}$ \cite{vi}, and then the terms inside the second $\exp(\cdot)$ in \eqref{opg} can be reformulated as
\begin{align}\label{refor}
& -\mathrm{Tr}\left(\sum_{l=1}^{L}\mathbf{P}_l^H\mathbf{P}_l\mathbf{G}\boldsymbol{\Xi}_l^{-1}\mathbf{G}^H\right.\left.-\mathbf{P}_l^H\mathbf{P}_l\boldsymbol{\Phi}_l \boldsymbol{\Xi}_l ^{-1}\mathbf{G}^H\right.\nonumber\\
&\left.-\mathbf{P}_l^H\mathbf{P}_l\mathbf{G}\boldsymbol{\Xi}_l^{-1}\boldsymbol{\Phi}_l^H\right)\nonumber\\
& = -\mathrm{Tr}\left(\sum_{l=1}^{L}\boldsymbol{\Xi}_l^{-1}\mathbf{G}^H\mathbf{P}_l^H\mathbf{P}_l\mathbf{G}\right.\nonumber\\
&\left.-\boldsymbol{\Xi}_l ^{-1}\boldsymbol{\Phi}_l^H \mathbf{P}_l^H\mathbf{P}_l \mathbf{G}-\boldsymbol{\Xi}_l^{-1}\mathbf{G}^H\mathbf{P}_l^H\mathbf{P}_l\boldsymbol{\Phi}_l\right)\nonumber\\
&=\sum_{l=1}^{L}-\mathrm{vec}(\mathbf{G}^H)^H\left(\mathbf{P}_l^H\mathbf{P}_l\otimes\boldsymbol{\Xi}_l^{-1}\right)\mathrm{vec}(\mathbf{G}^H)\nonumber\\
&+\mathrm{vec}(\boldsymbol{\Phi}_l^H)^H \left(\mathbf{P}_l^H\mathbf{P}_l\otimes\boldsymbol{\Xi}_l^{-1}\right)\mathrm{vec}(\mathbf{G}^H) \nonumber\\
&+\mathrm{vec}(\mathbf{G}^H)^H\left(\mathbf{P}_l^H\mathbf{P}_l\otimes\boldsymbol{\Xi}_l^{-1}\right)\mathrm{vec}(\boldsymbol{\Phi}_l^H).
\end{align}
Substituting \eqref{refor} into \eqref{opg}, we have
\begin{align}\label{appopg}
&Q^\dagger(\mathrm{vec}(\mathbf{G}^H))\!\propto\! \prod_{l=1}^{L}\mathcal{CN}_{NK}\left(\mathrm{vec}(\mathbf{G}^H)|\mathrm{vec}(\boldsymbol{\Phi}_l^H ),(\mathbf{P}_l^H\mathbf{P}_l)^{-1}\right.\nonumber\\
&\left.\otimes \boldsymbol{\Xi}_l\right)\times\mathcal{CN}_{NK}(\mathrm{vec}(\mathbf{G}^H)|\mathbf{0},(\mathbb{E}[\boldsymbol{\Gamma}]+\mathbb{E}[\mathbf{Z}])^{-1}).
\end{align}

Since \eqref{appopg} consists of the product of two Gaussian pdfs, after applying Lemma 1 in Appendix A, we have $
Q^\dagger(\mathrm{vec}(\mathbf{G}^H))= \mathcal{CN}_{NK}(\mathrm{vec}(\mathbf{G}^H)|\mathbf{u},\boldsymbol{\Omega})$
with
\begin{align}\label{gaur1}
\!\!\!\boldsymbol{\Omega}\!&=\!\left(\sum_{l=1}^{L}\mathbf{P}_l^H\mathbf{P}_l\otimes\boldsymbol{\Xi}_l^{-1}+\mathbb{E}[\boldsymbol{\Gamma}]+\mathbb{E}[\mathbf{Z}]\right)^{-1} \!\!\!\!\in \!\mathbb{C}^{NK \times NK},\\
\mathbf{u}&=\boldsymbol{\Omega}
\left(\sum_{l=1}^{L}\mathbf{P}_l^H\mathbf{P}_l\otimes\boldsymbol{\Xi}_l^{-1}\mathrm{vec}(\boldsymbol{\Phi}_l^H )\right)\in \mathbb{C}^{NK}.\label{gaur2}
\end{align}

Therefore, the pdf $Q^\dagger(\mathrm{vec}(\mathbf{G}))$ is a circularly symmetric complex vector-valued normal distribution with mean $\mathbf{u}$ and covariance $\boldsymbol{\Omega}$.
An intuitive interpretation of \eqref{gaur1} and \eqref{gaur2} are given as follows. The covariance matrix $\boldsymbol{\Omega}$ is updated by combining the posterior information from other factor matrices $\mathbf{X}_{l}^i, \forall l,i$, the
prior low-rank information $\mathbb{E}[\boldsymbol{\Gamma}]$, and prior element-wise sparsity information $\mathbb{E}[\mathbf{Z}]$. The tradeoff among these three terms is controlled by the expectation of the noise precision $\mathbb{E}[\beta]$ in $\boldsymbol{\Xi}_l$. In other words, the larger the current noise precision leads to more information from other factors than that from prior information.
On the other hand, the channel mean vector $\mathbf{u}$ is rotated by the covariance matrix $\boldsymbol{\Omega}$ to obtain the property of low-rank and element-wise sparsity.

Once the mean vector in \eqref{gaur2} is derived, the estimation of device-RIS reflected channel matrix $\mathbf{M}_G$ can be obtained. Note that the computation of $\boldsymbol{\Xi}_l$ in \eqref{opg} and \eqref{gaur2} is similar to \eqref{exi} and \eqref{ex2}, which will be introduced in the next section.

\subsection{Learning to Separate Devices: Inferring $Q(\mathbf{X}_{l}^i)$}
After obtaining the expression of $Q(\mathbf{G})$, we need to derive the update equation for each individual variational pdf $Q(\mathbf{X}_{l}^i)$, whose mean matrix is the estimated data message transmitted from the BS \cite{ste}.
After substituting \eqref{joint} into \eqref{kl3} and keeping the terms relevant to $\mathbf{X}_{l}^i$, the variational pdf $Q(\mathbf{X}_{l}^i)$ can be derived as follows:
\begin{figure*}
\begin{align}\label{opg3}
&Q^\dagger(\mathbf{X}_{l}^i)\propto \exp \left\{\mathbb{E}\left[-\beta\left \| \mathcal{Y}_l-\left[\!\!\left[\mathbf{X}_l^1,~\mathbf{X}_l^2,\cdots,\mathbf{X}_l^d, ~\mathbf{P}_l\mathbf{G}\right]\!\!\right]\right \|_F^2-\mathrm{Tr}(\boldsymbol{\Lambda}\mathbf{X}_{l}^{i,H}\mathbf{X}_{l})\right] \right\}\nonumber\\
&\propto  \exp  \left\{-\mathrm{Tr}\left(
 \mathbf{X}_{l}^i\right.\right.\left.\left.\underbrace{\mathbb{E}[\beta]\mathbb{E}\left[\left(\!\mathop \diamond \limits_{j=1,j\neq i}^{d}\mathbf{X}_{l}^j\mathop \diamond \limits \mathbf{P}_l\mathbf{G}\!\right)^T\!\left(\!\mathop \diamond \limits_{j=1,j\neq i}^{d}\mathbf{X}_{l}^j\mathop \diamond \limits \mathbf{P}_l\mathbf{G}\!\right)^*\right]\!+\!\mathbb{E}[\boldsymbol{\Lambda}]}_{(\boldsymbol{\Sigma }_l^i)^{-1} }\right.\right.\nonumber\\
&\left.\left.\times \! \mathbf{X}_{l}^{i,H}\!\!-\!\mathbf{X}_{l}^i(\boldsymbol{\Sigma }_l^i)^{-1}\![\underbrace{\mathbb{E}[\beta]\mathcal{Y}_l(i)\left(\mathop \diamond \limits \limits_{j=1,j\neq i}^{d}\mathbb{E}[\mathbf{X}_{l}^i]\diamond \mathbf{P}_l\mathbb{E}[\mathbf{G}] \right)^*\!\!\boldsymbol{\Sigma}_l^i}_{\mathbf{M}_l^i }]^H\right.\right.\left.\left.-\mathbf{M}_l^i (\boldsymbol{\Sigma }_l^i)^{-1}\mathbf{X}_{l}^{i,H}\right)\right\}.
\end{align}
\hrulefill
\end{figure*}
It can be concluded that the variational pdf $Q^\dagger(\mathbf{X}_{l}^i) (1 \leq l \leq L, 1 \leq i \leq d)$ is a circularly symmetric complex matrix normal distribution $\mathcal{CN}_{N\times K}(\mathbf{X}_{l}^i|\mathbf{M}_l^i,\mathbf{1}_{\tau_i}\otimes\boldsymbol{\Sigma}_l^i)$ with mean matrix $\mathbf{M}_l^i$ and covariance $\mathbf{1}_{\tau_i}\otimes\boldsymbol{\Sigma}_l^i$.

Once the matrix $\mathbf{M}_l^i$ in \eqref{opg3} at the top of next page is derived, the estimation of the transmitted message can be obtained.
All items in $\mathbf{M}_l^i$ can be calculated directly, except for $\boldsymbol{\Sigma }_l^i$. According to the property that $(\mathop \diamond\limits_{j=1}^{d}\mathbf{A}^j)^T(\mathop \diamond\limits_{j=1}^{d}\mathbf{A}^j)^*=\mathop \odot\limits_{j=1}^{d}\mathbf{A}^{j,T}\mathbf{A}^{j,*}$ \cite{chengpro}, the expectation of the coupled and individual factors terms in $(\boldsymbol{\Sigma}_l^i)^{-1}$ of \eqref{opg3} reduces to
\begin{align}\label{exi}
&\mathbb{E}\left[\left(\!\mathop \diamond \limits_{j=1,j\neq i}^{d}\mathbf{X}_{l}^j\mathop \diamond \limits \mathbf{P}_l\mathbf{G}\!\right)^T\!\left(\!\mathop \diamond \limits_{j=1,j\neq i}^{d}\mathbf{X}_{l}^j\mathop \diamond \limits \mathbf{P}_l\mathbf{G}\!\right)^*\right]\nonumber\\
&=\mathcal{D}\left[\mathop \odot\limits_{j=1,j\neq i}^{d}\mathbb{E}\left[\mathbf{X}_{l}^{j,T}\mathbf{X}_{l}^{j,*} \right]\odot\mathbb{E}\left[\mathbf{G}^T\mathbf{P}_l^T\mathbf{P}_l\mathbf{G}^*\right]\right],
\end{align}
where $\mathop \odot\limits_{j=1,j\neq i}^{d}\mathbf{A}^j=\mathbf{A}^d\odot\cdots\odot\mathbf{A}^{i+1}\odot\mathbf{A}^{i-1}\odot\mathbf{A}^1$ is the multiple Hadamard products.
The second term $\mathbb{E}\left[\mathbf{G}^T\mathbf{P}_l^T\mathbf{P}_l\mathbf{G}^*\right]$ on the right hand side (RHS) of \eqref{exi} can be derived by Lemma 2 in Appendix A and the first expectation is given by
\begin{align}\label{ex2}
\mathbb{E}\left[\mathbf{X}_{l}^{j,T}\mathbf{X}_{l}^{j,*} \right]=\mathbf{M}_l^{j,T}\mathbf{M}_l^{j,*}+{\tau_j}\boldsymbol{\Sigma}_l^j.
\end{align}

\subsection{Learning Hyper-Parameters: Inferring $Q(\boldsymbol{\xi})$, $Q(\boldsymbol{\eta})$, $Q(\boldsymbol{\gamma})$, and $Q(\beta)$}
Then, we infer the model parameters $\boldsymbol{\xi}(n,k)$ characterizing the element-wise sparsity of the channel. After plugging the postulated joint pdf \eqref{joint} into \eqref{kl3} and only remaining the terms about $\boldsymbol{\xi}(n,k)$, we have $
Q^\dagger({\{\boldsymbol{\xi}(n,k)\}_{n=1}^N}_{k=1}^K)=\prod_{n=1}^{N}\prod_{k=1}^{K}Q^\dagger(\boldsymbol{\xi}(n,k))$
with
\begin{align}\label{joxi1}
&Q^\dagger(\boldsymbol{\xi}(n,k))\propto  \exp\{-\boldsymbol{\xi}(n,k)
(\underbrace{\mathbb{E}\left[\mathbf{G}(n,k)^*\mathbf{G}(n,k)\right]+\delta}_{a_{\boldsymbol{\xi}(n,k)}})\nonumber\\
&+(\underbrace {N+\delta}_{b_{\boldsymbol{\xi}}}-1)\ln\boldsymbol{\xi}(n,k)\}.
\end{align}
Since the optimal $Q^\dagger(\boldsymbol{\xi}(n,k))$ in \eqref{joxi1} obeys a gamma distribution of the form
$Q^\dagger (\boldsymbol{\xi}(n,k))=\mathrm{gamma}(\boldsymbol{\xi}(n,k)|a_{\boldsymbol{\xi}(n,k)},b_{\boldsymbol{\xi}})$, the mean value of $\boldsymbol{\xi}(n,k)$ can be updated as $\mathbb{E}[\boldsymbol{\xi}(n,k)]=b_{\boldsymbol{\xi}}/a_{\boldsymbol{\xi}(n,k)}$.
Using the correlation property of the normal distribution derived in \eqref{mean}, the expectation in $a_{\boldsymbol{\xi}(n,k)}$ can be calculated as
\begin{equation}\label{al}
\!\!\mathbb{E}\left[\mathbf{G}(n,k)^*\mathbf{G}(n,k)\right]\!=\!\mathbf{M}_{G}(n,k)^*\mathbf{M}_{G}(n,k)\!+\!\boldsymbol{\Omega}_{n,n}^b(k,k).\!\!\!\!\!
\end{equation}

Next, we infer the model parameters $\eta_{k}$ and $\gamma_{k}$ characterizing the low rank of the tensor data $\mathcal{Y}_l$. Combining \eqref{joint} and \eqref{kl3} and removing the terms independent of $\eta_{k}$, we obtain
\begin{align}\label{joeta}
&Q^\dagger(\{\eta_k\}_{k=1}^K)\propto  \exp\{\mathbb{E}[
-\mathrm{Tr}(\boldsymbol{\Lambda}\mathbf{G}^H\mathbf{G})-\sum_{k=1}^{K}\delta\eta_k \nonumber\\
& +\left(N+\delta-1\right)\sum_{k=1}^{K}\ln{\eta_k}]\},
\end{align}
which can be reformulated as $
Q^\dagger(\{\eta_k\}_{k=1}^K)=\prod_{k=1}^{K}Q^\dagger(\eta_k)$
with
\begin{align}\label{joeta2}
\!\!\!Q^\dagger(\eta_k)\!\propto \! \exp \{-\eta_{k}
\underbrace{\mathbb{E}\left[\mathbf{G}(:,k)^H\mathbf{G}(:,k)
+\delta\right]}_{a_{\eta_k}}\!+\!(\underbrace{N\!+\!\delta}_{b_{\eta}}-1)\ln\eta_{k}\}.
\end{align}
From \eqref{joeta2}, we know that the optimal $Q^\dagger(\eta_k)$ obeys a gamma distribution of the form $\mathrm{gamma}(\eta_k|a_{\eta_k},b_{\eta})$. Taking the expectation in $a_{\eta_k}$ and utilizing \eqref{mean} in Lemma 2 in Appendix A, we have
\begin{align}\label{eeta1}
&a_{\eta_k}=\mathbf{M}_{G}(:,k)^H\mathbf{M}_{G}(:,k)\!+\!\{\mathop\sum \limits_{n=1}^{N}\boldsymbol{\Omega}_{n,n}^b\}(k,k)+\delta.
\end{align}
Thus the expectation of parameter $\eta_k$ is given by $\mathbb{E}(\eta_k)=b_{\eta}/a_{\eta_k}$ with $b_{\eta}=N+\eta$. In a similar way, the optimal $Q^\dagger(\gamma_k)$ is found to be a gamma distribution of the form $\mathrm{gamma}(\gamma_k|a_{\gamma_k},b_{\gamma})$ with
$
a_{\gamma_k}=\mathbb{E}\left[\sum_{l=1}^{L}\sum_{i=1}^{d}\mathbf{X}_{l}^i(:,k)^{H}\mathbf{X}_{l}^i(:,k)
+\delta\right]$
and $b_{\gamma}=L\sum_{i=1}^{d}\tau_i+\delta$. Herein, $\mathbb{E}\left[\mathbf{X}_{l}^i(:,k)^{H}\mathbf{X}_{l}^i(:,k)\right]=\mathbf{M}_{l}(:,k)^{i,H}\mathbf{M}_{l}(:,k)^{i}+{\tau_i}\boldsymbol{\Sigma}_{l}^i(k,k)
$. Then we can calculate the expectation of parameter $\gamma_k$ by $\mathbb{E}(\gamma_k)=b_{\gamma}/a_{\gamma_k}$.

Finally, the posterior distribution of $\beta$ is updated by the following equation:
\begin{align}\label{jbe}
&Q^\dagger(\beta)\propto (\underbrace{(\Pi_{i=1}^{d+1}\tau_iM)L+\delta}_{b_{\beta}}-1)\ln\beta\nonumber\\
&-\beta\underbrace{\sum_{l=1}^{L}\mathbb{E}\left[\left \| \mathcal{Y}_l-\left[\!\!\left[\mathbf{X}_l^1,~\mathbf{X}_l^2,\cdots,\mathbf{X}_l^d, ~\mathbf{P}_l\mathbf{G}\right]\!\!\right]\right \|_F^2+\delta\right]}_{a_{\beta}}.\!\!\!
\end{align}
By comparing \eqref{jbe} to the functional form of gamma distribution, we have $Q^\dagger(\beta)=\mathrm{gamma}(\beta|a_{\beta},b_{\beta})$. Herein, $b_{\beta}$ in \eqref{jbe}
is related to the number of observations and $a_{\beta}$  approximates the residual of model fitting measured by the squared Frobenius norm on observed entries.

For calculating $a_{\beta}$ in \eqref{jbe}, we apply the tensor unfolding and then expand the Frobenius norm as follows
\begin{align}\label{exp}
&\mathbb{E}\left[\left \| \mathcal{Y}_l-\left[\!\!\left[\mathbf{X}_l^1,~\mathbf{X}_l^2,\cdots,\mathbf{X}_l^d, ~\mathbf{P}_l\mathbf{G}\right]\!\!\right]\right \|_F^2\right]\nonumber\\
&=\mathrm{Tr}\left(\mathop \odot \limits _{i=1}^{d}\left(\mathbf{M}_l^{i,H}\mathbf{M}_l^{i,*}+{\tau_i}\boldsymbol{\Sigma}_l^i\right)^H\times \left(\mathbf{M}_G^H\mathbf{P}_l^H\mathbf{P}_l\mathbf{M}_G\right.\right.\nonumber\\
&\left.\left.+\mathop\sum \limits_{i=1}^{N}\mathop\sum \limits_{j=1}^{N}\{\mathbf{P}_l^H\mathbf{P}_l\}(i,j)\boldsymbol{\Omega}_{i,j}^{b}\right)+\left\|\mathcal{Y}_l(d+1)\right\|_F^2\right.\nonumber\\
&\left.-\mathbf{P}_l\mathbf{M}_G
\left(\mathop \diamond \limits _{i=1}^{d}\mathbf{M}_l^{i}\right)^T\mathcal{Y}_l(d+1)^H\right.\nonumber\\
&\left.-\mathcal{Y}_l(d+1)\left(\mathop \diamond \limits _{i=1}^{d}\mathbf{M}_l^{i}\right)^*\mathbf{M}_G^H\mathbf{P}_l^H\right),
\end{align}
where the RHS of \eqref{exp} is obtained by utilizing equation \eqref{mean1} in Lemma 2 in Appendix A and the result in \eqref{ex2}.

\begin{remark}{
(Learning of Active Device Number): During the iterations, the mean of the model complexity parameters, i.e., $\mathbb{E}(\eta_k)$ and $\mathbb{E}(\gamma_k)$ are learned by equations
$\mathbb{E}(\eta_k^{t+1})=b_{\eta}/a_{\eta_k}^{t+1}$ and $\mathbb{E}(\gamma_k^{t+1})=b_{\gamma}/a_{\gamma_k}^{t+1}$, respectively, together with those of other parameters. Since $\mathbb{E}(\eta_k^{-1})$ and $\mathbb{E}(\gamma_k^{-1})$ have physical interpretations as the power of each column in $\mathbf{G}$ and $\mathbf{X}_l^i$, if some $\mathbb{E}(\eta_k)$ and $\mathbb{E}(\gamma_k)$ are large enough, it indicates that their corresponding columns in $\mathbf{G}$ and $\mathbf{X}_l^i$ can be safely pruned out.
Then, the number of non-zero columns in each factor matrix equals to the number of active devices.}
\end{remark}

\subsection{Algorithm Summary}
Since the statistics of each variational pdf rely on other variational pdfs. Therefore, all parameters of variational pdfs need to be updated alternatingly. For clarity, the pseudo-code of the resulting algorithm is outlined in \textbf{Algorithm 1}.
\begin{breakablealgorithm}
\caption{Coupled Tensor-Based Automatic Detection (CTAD) Algorithm}
\label{alg1}
\begin{algorithmic}[t]
\State  \textbf{Input}: $\{\mathcal{Y}_l\}_{l=1}^L$ and total iterations $T$
\State \textbf{Initialization}: $\mathbf{M}_G^0$, ${\{\mathbf{M}_l^{i,0}\}_{l=1}^{L}}_{i=1}^{d}$, $\boldsymbol{\Sigma}_{G}^0$, ${\{\boldsymbol{\Sigma}_l^{i,t}\}_{l=1}^{L}}_{i=1}^{d}$, $a_{\beta}^0$, $\{a_{\eta_k}^0,a_{\gamma_k}^0\}_{k=1}^K$, and ${\{a_{{\boldsymbol{\xi}}_{n,k}}^0\}_{n=1}^{N}}_{k=1}^{K}$
\For{$t=1 : T$}
\State {\bf{Updates the parameters of $Q(\mathbf{G})^{t+1}$}}:
{\setlength\abovedisplayskip{0.5pt}
\setlength\belowdisplayskip{0.5pt}
\begin{align}\label{al2}
\boldsymbol{\Xi}_l^{t+1}&=\left( \frac{a_{\beta}^t}{b_{\beta}}\mathcal{D}\left[\mathop \odot\limits_{i=1}^{d}\left((\mathbf{M}_l^{i,t})^H\mathbf{M}_l^{i,t}+{\tau_i}\boldsymbol{\Sigma}_l^{i,t}\right)^*\right]\right)^{-1},
\end{align}}
\begin{align}
&{\boldsymbol{\Omega}}^{t+1}=\left(\sum_{l=1}^{L}\mathbf{P}_l^H\mathbf{P}_l\otimes\left(\boldsymbol{\Xi}_l^{t+1}\right)^{-1}\right.\nonumber\\
&\left.+\text{diag}\left(\mathbf{1}_{N}\otimes\left[ \frac{b_{\eta}}{a_{\eta_1}^t },\cdots,\frac{b_{\eta}}{a_{\eta_K}^t }\right]\right)\right.+\text{diag}\left( \frac{b_{\boldsymbol{\xi}}}{a_{\boldsymbol{\xi}_{1,1}}^t},\right.\nonumber\\
&\left.\left.\cdots,\frac{b_{\boldsymbol{\xi}}}{a_{\boldsymbol{\xi}_{1,K}}^t},\cdots,\frac{b_{\boldsymbol{\xi}}}{a_{\boldsymbol{\xi}_{N,1}}^t},\cdots,\frac{b_{\boldsymbol{\xi}}}{a_{\boldsymbol{\xi}_{N,K}}^t}\right)\right)^{-1} ,\label{al1}
\end{align}
\begin{align}
&\mathbf{u}^{t+1}={\boldsymbol{\Omega}}^{t+1}\mathop \sum \limits_{l=1}^{L}\mathbf{P}_l^H\mathbf{P}_l\otimes(\boldsymbol{\Xi}_l^{t+1
})^{-1}\nonumber\\
&\mathrm{vec}\!\left(\!\left(\!\frac{a_{\beta}^t}{b_{\beta}}(\mathbf{P}_l^H\mathbf{P}_l)^{-1}\!\mathbf{P}_l^{H}\mathcal{Y}_l(d+1)\left(\mathop \diamond \limits_{i=1}^{d}\mathbf{M}_{l}^{i,t}\right)^*\boldsymbol{\Xi}_l^{t+1
}\!\right)^H\!\right)\!,\!\!\!\!\label{uut}
\end{align}
\begin{align}\label{ut1}
\mathbf{M}_G &=\mathrm{reshape}(\mathbf{u},N \times K).
\end{align}

\State {\bf{Updates the parameters of $Q(\mathbf{X}_l^i)^{t+1}$}}:
\begin{align}\label{x1}
&\boldsymbol{\Sigma}_l^{i,t+1}\!=\!\!\left(\frac{a_{\beta}^t}{b_{\beta}}\mathcal{D}\!\left[\!\mathop \odot\limits_{j=1,j\neq i}^{d}\!\left(\!(\mathbf{M}_l^{j,t})^H\mathbf{M}_l^{j,t}\!+\!{\tau_j}\boldsymbol{\Sigma}_l^{j,t}\right)^*
\!\right.\right.\!\!\!\!\!\!\!\!\!\!\!\!\nonumber\\
&\left.\left.\!\odot\!\left(\!\mathbf{M}_G^{t,H}\mathbf{P}_l^H\mathbf{P}_l\mathbf{M}_G^t+\mathop\sum \limits_{i=1}^{N}\mathop\sum \limits_{j=1}^{N}\{\mathbf{P}_l^H\mathbf{P}_l\}(i,j)\boldsymbol{\Omega}_{i,j}^{b,t}\!\right)^*\right]\right.\nonumber\\
&\left.+\text{diag}\left(\frac{b_{\gamma}}{a_{\gamma_1}^t },\cdots,\frac{b_{\gamma}}{a_{\gamma_K}^t }\right)\right)^{-1},
\end{align}
\begin{align}\label{x2}
\mathbf{M}_l^{i,t}&=\frac{a_{\beta}^t}{b_{\beta}}\mathcal{Y}_l(i)\left(\mathop \diamond \limits \limits_{j=1,j\neq i}^{d}\mathbf{M}_{l}^{i,t}\diamond \mathbf{P}_l\mathbf{M}_G^t\right)^*\boldsymbol{\Sigma}_l^{i,t+1}.
\end{align}
\State {\bf{Updates the parameters of $Q(\boldsymbol{\xi}(n,k))^{t+1}$}}:
{\setlength\abovedisplayskip{0.5pt}
\setlength\belowdisplayskip{0.5pt}
\begin{align}\label{alx}
\!\!\!\!a_{\boldsymbol{\xi}(n,k)}^{t+1}=\mathbf{M}_{G}^{t+1,*}(n,k)\mathbf{M}_{G}^{t+1}(n,k)+\boldsymbol{\Omega}_{n,n}^{b,t+1}(k,k).
\end{align}}
\State {\bf{Updates the parameters of $Q(\eta_{k})^{t+1}$}}:
{\setlength\abovedisplayskip{0.5pt}
\setlength\belowdisplayskip{0.5pt}
\begin{align}\label{alx1}
\!\!\!a_{\eta_k}^{t+1}&\!=\!(\mathbf{M}_{G}^t(:,k)^{H}\mathbf{M}_{G}(:,k)^t
\!+\![\mathop\sum \limits_{n=1}^{N}\boldsymbol{\Omega}_{n,n}^{b,t+1}](k,k))
\!+\!\delta.
\end{align}}
\State {\bf{Updates the parameters of $Q(\gamma_{k})^{t+1}$}}:
{\setlength\abovedisplayskip{0.5pt}
\setlength\belowdisplayskip{0.5pt}
\begin{align}\label{alx3}
a_{\gamma_k}^{t+1}&=\sum_{l=1}^{L}\sum_{i=1}^{d}(\mathbf{M}_{l}^{i,t}(:,k))^H\mathbf{M}_{l}^{i,t}(:,k)+{\tau_i}\boldsymbol{\Sigma}_{l}^{i,t}(k,k) +\delta.
\end{align}}
\State {\bf{Updates the parameters of $Q(\boldsymbol{\beta})^{t+1}$}}:
{\setlength\abovedisplayskip{0.5pt}
\setlength\belowdisplayskip{0.5pt}
\begin{align}\label{alx5}
a_{\beta}^{t+1}=\sum_{l=1}^{L}\pounds_l^{t+1}+\delta,
\end{align}}
~~~where $\pounds_l^{t+1}$ is calculated by \eqref{exp}.
\EndFor
\State \textbf{Output}: $\mathbf{M}_G^{t+1}$ and ${\{\mathbf{M}_l^{i,t+1}\}_{l=1}^L}_{k=1}^K$
\end{algorithmic}
\end{breakablealgorithm}

We now provide the details of Algorithm 1: Eq. \eqref{al2} denotes the variance of factor matrix $\mathbf{G}$, and Eqs. \eqref{al1} and \eqref{uut} denote the variance and mean of vector $\mathrm{vec}(\mathbf{G})$, respectively, which is derived in Section IV-C. Eq. \eqref{ut1} denotes the mean in the matrix form of factor matrix $\mathbf{G}$, which is reshaped from vector $\mathbf{u}$. Eqs. \eqref{x1} and \eqref{x2} denote the variance and mean of factor matrix $\mathbf{X}$, respectively, which are derived in Section IV-D. Eqs. \eqref{alx}-\eqref{alx5} denote the hyper-parameters of $\boldsymbol{\xi}$, $\boldsymbol{\eta}$, $\boldsymbol{\gamma}$, and $\beta$, respectively, which are derived in Section IV-E.
After applying \textbf{Algorithm 1}, we can obtain ${\{\mathbf{M}_l^{i,t+1}\}_{l=1}^L}_{k=1}^K$ and $\mathbf{M}_G^{t+1}$, which correspond to the estimation of data $\mathbf{x}_{i,k,l}, \forall i, k, l$, and the channel sparsity profile $\boldsymbol{\lambda}_k, \forall k$. Note that scalar indeterminacy can be resolved by designing each sub-constellation $\wp_i$ based on a Grassmannian codebook \cite{grcode2}.
\subsection{Performance Analysis}
To gain further insights from the above proposed algorithm, this subsection analyzes its performance from the perspective of computational complexity and the convergence property.
\subsubsection{Computational Complexity}
For each iteration, the computational complexity is dominated by updating each factor matrix and mainly arises from the matrix multiplication, which is in the order of $\mathcal{O}(L (\mathop \sum\limits_{i=1}^{d}\tau_i+M)K^3+Ld\mathop \prod_{i=1}^{d}M\tau_iK^2+LN^2M)$. It can be seen that the complexity of the proposed CTAD algorithm scales linearly with the subblocks but only polynomially with the potential number of active devices. In contrast, the computational complexity of the active device separation algorithm in \cite{shi} scales polynomially with the total number of devices, i.e., $\mathcal{O}(\bar{K}^4 )$. Thus, the proposed CTAD algorithm is computational efficiency.

\subsubsection{Convergence Property}
The functional minimization of the KL divergence in \eqref{kl} is convex with respect to a single variational pdf $Q(\Theta_j)$ when the others $\{Q(\Theta_j)\}_{i\neq j}$ are fixed. This guarantees monotonic decrease of the KL divergence in \eqref{kl} over iterations. In addition, the proposed CTAD algorithm is an instance of the block coordinate descent optimization strategy over the functional space. Therefore, the CTAD algorithm is able to decrease the objective in every iteration. Since the proposed algorithm is developed under the mean-field variational inference framework \cite{vi}, the limit point of the proposed algorithm is guaranteed to converge to a stationary point of the objective function in \eqref{kl}.
\subsubsection{Identifiability Property}
Uniqueness is a key characteristic of the proposed tensor decomposition problem.  If $LM\geq 2\varpi$ and $2K_a+d\leq \min\{\kappa_1,\kappa_2,\cdots,\kappa_L\}$ with $\kappa_l=\sum_{i=1}^{d}r_{\mathbf{X}_l^i}+\min(M,r_\mathbf{G})$, then $\mathbf{X}_l^i$ and $\mathbf{G}$ are almost surely identifiable from the received data $\mathcal{Y}_l, l=1,2,\cdots,L$.
Herein, $r_{\mathbf{X}_l^i}$ is the Kruskal-rank of each factor matrix $\mathbf{X}_l^i$ and $\varpi $ is an upper bound on the number of nonzero elements in each column of $\mathbf{G}$.

\emph{Remark 2}:  In this paper, we consider a scenario of Rayleigh fading channels without line-of-sight (LoS) paths between the devices and the RIS. In fact, due to the massive number of devices in a cell, there may exist LoS paths between the devices and the RIS. In such a scenario, the proposed tensor-based URA scheme is also applicable by redesigning the  prior for matrix $\mathbf{G}$ in \eqref{pen3}.
Hence, the proposed URA scheme has a wide application.

\section{Numerical Results}
In this section, we present extensive simulation results to validate the effectiveness of the proposed CTAD algorithm in 6G wireless networks.
For the URA scheme, we consider the case that every active device embeds its ID in the payload, in addition to the $B = 360$ information bits \cite{unsourced}. It is assumed that there are total of $\bar{K}=4096$ devices in the network, therefore, $B_{\mathrm{ID}} = \log_{2}(K) = 12$ bits are required to encode the device ID.
Hence, each device transmits a total of $B_{\mathrm{tot}} = B_{\mathrm{ID}}+B = 372$ bits.
At the BS, the active devices are identified based on the ID embedded in the decoded payload. The active device separation error is evaluated by the packet error rate (PER) metric defined as $\mathrm{PER}=\frac{\bar{K}_a}{{K_a}}$, where $\bar{K_a}$ is the number of incorrectly decoded messages. If multiple messages with
the same device ID are decoded, they are discarded and
counted as an error. We assume that the pilot sequences obey an independent and identically distributed (i.i.d.) Gaussian distribution with zero mean and unit variance. Every $\epsilon_i^k$ of the BS-device channel is drawn from the complex Gaussian distributed random variables with zero mean and unit variance. The distance $d_{\mathrm{DR},k}$ from the $k$th device to the RIS is randomly generated from $500$ m to $1000$ m and from the RIS to the BS is $100$ m. Note that the models in \eqref{U} and \eqref{angu} are 3D channel models. We draw the central azimuth and elevation AoD from the RIS (AoA to the RIS) of each cluster uniformly over $[-180^{\circ},180^{\circ}]$ and $[-90^{\circ},90^{\circ}]$, respectively; draw the central azimuth AoA at the BS of each cluster uniformly over $[-90^{\circ},90^{\circ}]$. The angular spread of each subpath is set as $15^{\circ}$. The over-complete bases $\boldsymbol{\nu}$ and $\boldsymbol{\varsigma}$ in \eqref{angu3} are uniform sampling grids covering $[-1,1]$. The channel estimation accuracy is evaluated in terms of normalized
mean square error (NMSE) given by
$\mathrm{NMSE}=\frac{1}{T_r}\mathop \sum \limits_{i=1}^{T_r}\frac{\left \|{\mathbf{G}}^i-\hat{\mathbf{G}}^i\right \|_F^2}{\left \|{\mathbf{G}}^i\right \|_F^2}$, where $T_r$ is the number of Monte Carlo runs and $\hat{\mathbf{G}}^i$ is the device-RIS channel estimated at the $i$-th run. The tree-based decoder \cite{unsourced} is employed in the simulation, where the $B$-bit message of each device is divided into $L$ subblocks of size $B_1,\cdots,B_{L}$ satisfying the following conditions: $\sum_{l}B_l=B$, $B_1=R$, and $B_l<R$ for all $i=2, \cdots, L$.
Herein, all subblocks $i=2,\cdots,L$ are augmented to size $R$ by appending the parity bits which are set to $p_{b}(0,168,270)$. Unless stated otherwise, we set $R=270$, $N_1'/N_1=N_2'/N_2=1.5$, $\tau_p=N$, $\delta=10^{-6}$, $d=2$, $\tau_1=80$, and $\tau_2=80$.

For comparison, we consider a baseline scheme which employs a two-stage estimator for addressing the optimization problem in \eqref{eT}: First, the alternating least square (ALS) algorithm \cite{tenrank} is applied to fit the model in \eqref{eT} and $\bar{\mathbf{G}}_l=\mathbf{P}_l\mathbf{G}$ is regarded unknown to estimate. As for the second stage, we compute the minimum $l_1$-norm solutions for the sparse factor matrices $\mathbf{G}$ from the estimation of ${\bar{\mathbf{G}}}_l$ and use the existing convex
optimization solver \cite{convex} to solve this least absolute shrinkage and selection operator (LASSO) problem. Since the number of active devices $K_a$ is unknown, we set the initial upper bounds of $K_a$ as $K$ in the two-stage estimator.
\begin{figure}[t]
  \centering
\includegraphics [width=85mm] {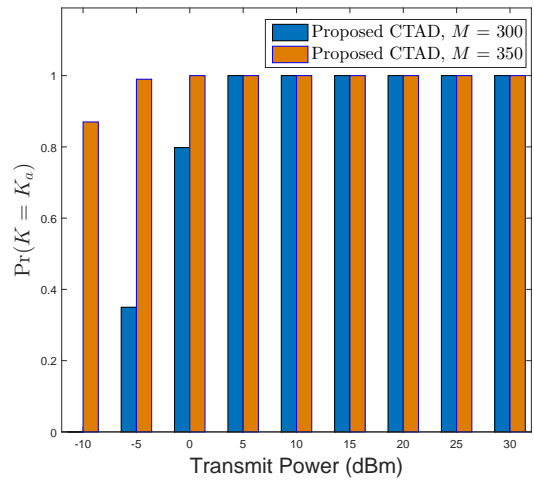}
\caption{The estimation accuracy of the number of active devices for different transmit power with $K_a=50$, $N =500$, $\bar{K}=4096$.}
\label{rank}
\end{figure}

\begin{figure}[t]
  \centering
\includegraphics [width=85mm] {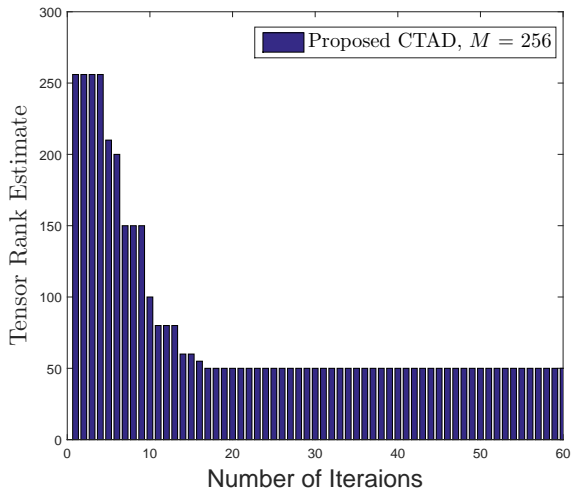}
\caption{The tensor rank estimate versus number of iterations with $K_a=50$, $N = 400$, transmit power $=15$ dBm, and $\bar{K}=4096$.}
\label{rankiter}
\end{figure}
When measuring the rank quality of estimation, we consider the mean value of the rank
estimation and probability of the successful recovery in the form of $\mathrm{Pr}(K = K_a)$. Each simulation is repeated 200 times to obtain the averaged result. Fig. \ref{rank} suggests that for $M = 300$ and the data transmit
power $\geq 5 $ dBm, the proposed CTAD algorithm can recover the true tensor rank with probability 1 in the considered simulation setting. This shows the high accuracy and robustness of the proposed algorithm when the noise power is moderate.
For $M = 350$, a cut-off point for accurate rank estimation is relaxed to transmit
power $\geq -5 $ dBm, which indicates that the success rate of rank estimation is increased as the antenna number $M$ at the BS grows in the lower transmit power region.
This is due to the fact that $\mathbf{P}_l$ is equivalent to the measurement matrix in compressed sensing and the increasing number of BS antennas results in the increment of measurement length for a better observation. Fig. \ref{rankiter} shows the mean value of estimated rank along with the number of iterations. It can be seen that only a few iterations are required for the convergence of the proposed algorithm on average. In particular, the proposed algorithm has gradually recovered the number of active devices, while the estimated rank remains stable. This is because, as illustrated in Remark 1, the regularization parameters of these active devices are kept over the next iteration, while the values of the other devices are pruned out by thresholding.
\begin{figure}[t]
\centering
\includegraphics[width=85mm]{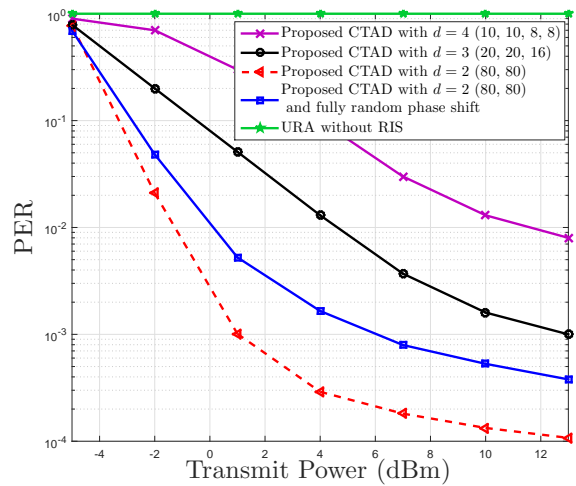}
\caption{PER versus different tensor sizes $d$ and $\tau_i$ with $M= 350$, $N=500$, $L=3$, $K_a=15$, and $\bar{K}=4096$.}
\label{PERdNum}
\end{figure}

\begin{figure}[t]
\centering
\includegraphics[width=85mm]{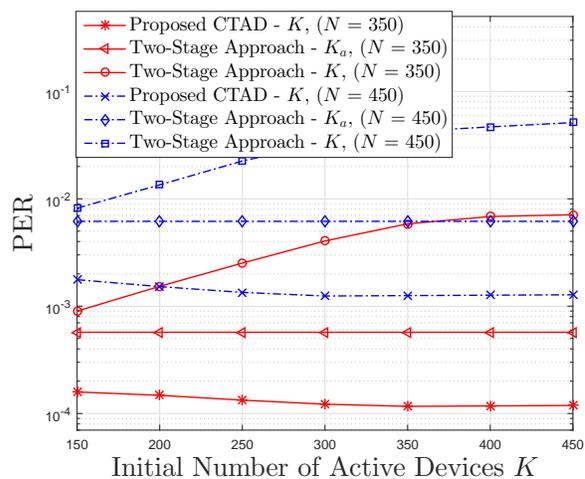}
\caption{PER versus different $K$ with $K_a=50$, $M=256$, transmit power $=10$ dBm, and $\bar{K}=4096$.}
\label{PERBound}
\end{figure}

\begin{figure}[t]
\centering
\includegraphics[width=85mm]{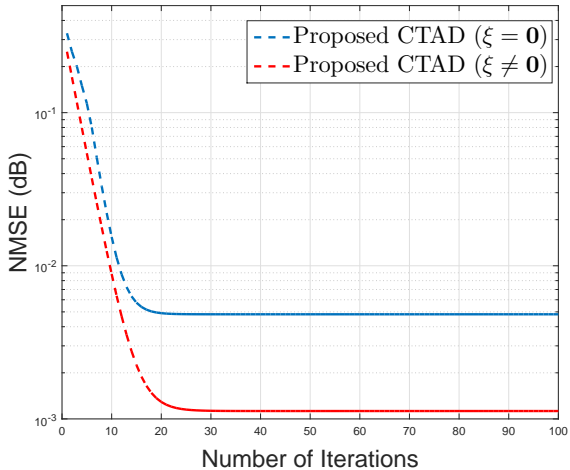}
\caption{The convergence performance with $K_a=50$, $M=350$, $N=500$, transmit power $=15$ dBm, and $\bar{K}=4096$.}
\label{convergence}
\end{figure}

Fig. \ref{PERdNum} plots the PER against transmit power for varying
tensor sizes in an RIS-aided massive unsourced access system. {It can be seen that URA scheme in \cite{newT} cannot recover any
active device without the RIS under unfavorable propagation conditions. The reason is that we assume that the direct link between the BS and the devices is completely blocked and unavailable, and hence the BS cannot receive the signal sending from the devices.
However, this issue can be resolved by deploying an RIS in URA networks. Moreover, we can observe that the optimized/random combined phase shift achieves significant performance gains over the proposed CTAD with fully random phase shifts. Again, this confirms the effectiveness of the optimized/random combined beamforming design for RIS-aided URA  communications.
Note that the PER of the CTAD algorithm with $d=4$ and $d=3$ are higher than that of the CTAD algorithm with $d=2$. This can be explained through the number of DoF per active device. Since a variable in Grassmannian of lines in dimension $\tau_i$ has $\tau_i-1$ DoF \cite{grcode2}, the sum-DoF of the active devices in the considered model can be defined as $\mathrm{DoF}(K_a)=K_a\sum \limits_{i=1}^{d}(\tau_i-1)$.
As such, for a fixed number of active devices $K_a$, we can see that the available DoF of $d=4$ and $d=3$ are lower than that of $d=2$, which makes $d=2$ more suitable for RIS-aided URA communications.

In Fig. \ref{PERBound}, we provide the PER performance curves of the proposed CTAD algorithm under different initial upper bounds $k$ of $K_a$ available, where the genie-aided two-stage approach with exact $K_a$ and the two-stage approach with incorrect active device numbers $K$ are set as benchmarks. First, it can be observed that the proposed CTAD algorithm can offer the best PER results due to its superior capabilities in determining the exact tensor rank, i.e., $K_a$. In contrast, the two-stage approach with $K$ overfits the noise heavily and the performance degrades severely in channel estimation. Second, the genie-aided two-stage approach with $K_a$ still performs worse than the proposed CTAD algorithm. The reason is that the proposed CTAD algorithm exploits the prior statistical
distributions of the sparse channel, which is ignored by
the two-stage approach. Third, increasing the number of reflecting elements causes a PER degradation, as the dimension of channel estimates improves with an increment in the reflecting elements.

Then, we show the convergence behaviour of the proposed CTAD algorithm in Fig. \ref{convergence} by setting $\boldsymbol{\xi}=\mathbf{0}$ and $\boldsymbol{\xi} \neq\mathbf{0}$, respectively. It is observed that the NMSE of the proposed
algorithm with $\boldsymbol{\xi} \neq\mathbf{0}$ decreases
quickly with the increment of the number of iterations and the algorithm
converges to stable values in less than 25 iterations on average. Also, Fig. \ref{convergence} verifies that the proposed algorithm significantly outperforms the algorithm that does not explore the element-wise sparsity ($\boldsymbol{\xi} =\mathbf{0}$).
The reason is that when $\boldsymbol{\xi}=\mathbf{0}$, although the algorithm searches for a low-rank structure by automatically adjusting parameters $\eta_k$ and $\gamma_k$, it is not
able to explore element sparsity as each column of the
channel matrix shares the same hyperparameter. Hence, the
estimation accuracy of model parameters is unsatisfactory, which in turn degrades its ability to decompose tensors. In contrast, the proposed CTAD algorithm with $\boldsymbol{\xi} \neq\mathbf{0}$ infers
the dynamics of the network by using the power of columns of channel.
\begin{figure}[t]
\centering
\includegraphics[width=85mm]{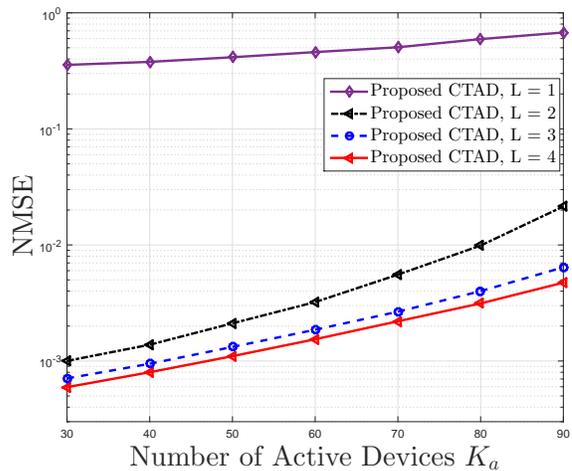}
\caption{NMSE versus $L$ with $M= 256$, $N=350$, transmit power $=15$ dBm, and $\bar{K}=4096$.}
\label{nmseactivity}
\end{figure}

\begin{figure}[t]
\centering
\includegraphics[width=85mm]{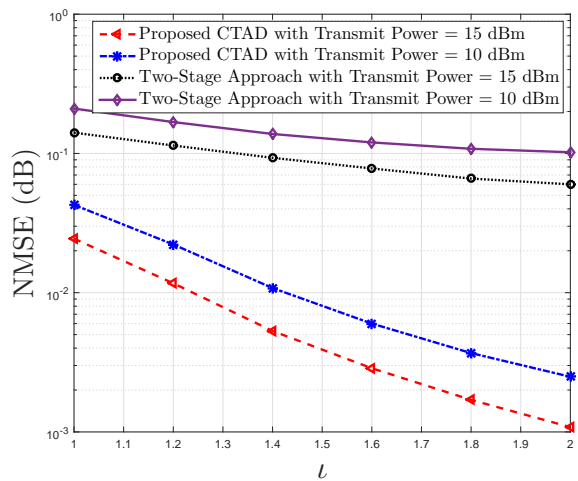}
\caption{The NMSE versus sampling resolution $\iota$ with $M= 250$, $N=500$, $K_a=50$, and $\bar{K}=4096$.}
\label{sampling}
\end{figure}

\begin{figure}[t]
\centering
\includegraphics[width=85mm]{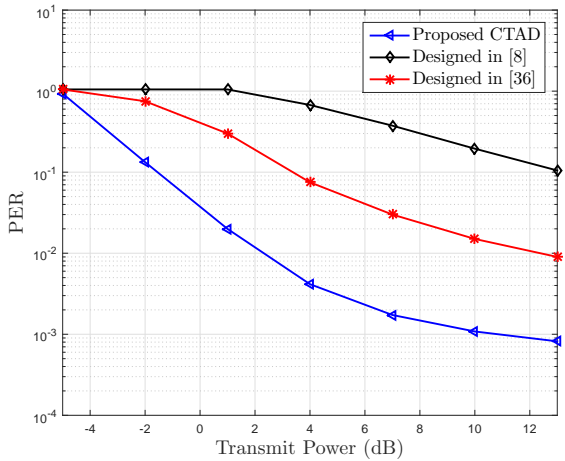}
\caption{PER versus transmit power under the case where the direct channel link is not completely blocked.}
\label{directweak}
\end{figure}

Next, we focus on the NMSE performance of the proposed CTAD algorithm with different numbers of subblocks $L$ and a fixed rate $B/\tau$ in Fig. \ref{nmseactivity}. As can be seen from the obtained results, utilizing a single subblock $L=1$ cannot accurately estimate the channel, since $\mathbf{G}$ does not admit a unique solution from the compressed model $\mathbf{P}_l\mathbf{G}$. While the proposed CTAD algorithm is quite accurate for both $L=3, 4$ and the performance degrades with decreasing $L$. From this simulation result, we can see that the coupled tensor factorization criterion in problem \eqref{eT} is critical to the
reconstruction of the channel, since the shared parameter $\mathbf{G}$ in the $L$ fitting terms serves as an anchor to fix the permutation and scaling ambiguities. As the difference in PER performances of $L=3$ and $L=4$ is relatively smaller, $L$ is set to 3 in the rest of the simulations for unveiling the full potential of the CTAD algorithm under various system settings.

In Fig. \ref{sampling}, we study the effect of the grid lengths $N_1'$ and $N_2'$. We adopt $\iota=N_1'/N_1=N_2'/N_2$ to represent the ratio between the grid length and the number of reflecting elements. It is seen that the NMSE decreases as $\iota$ increases, since increasing the sampling grid length leads
to a higher angular resolution and hence results in a sparser $\mathbf{G}$.
The proposed algorithm substantially decreases the NMSE compared with the two-stage approach, and the performance gaps among the two algorithms decrease as the ratio decreases.
This is because the NMSE performance is limited by the
sparsity of the channel.
Besides, the performances of all estimation
schemes increase as the transmit power grows due to a better received signal quality at the BS.

Finally, we compare the proposed CTAD algorithm and the URA schemes designed in \cite{unsourced} and \cite{newT}. In Fig. \ref{directweak}, we consider a scenario that the direct channel link between the BS and devices is not completely blocked such that a weak signal can reach the BS through the direct link. It is assumed that the direct channel link obeys an i.i.d. Gaussian distribution with zero mean and $0.3$ variance.
Without the aid of RIS, the URA schemes designed in \cite{unsourced} and \cite{newT} perform active device detection only with weak direct link signals. Accordingly, the proposed CTAD algorithm performs active device separation only based on IRS-associated signals. It can be seen that the PER performance of the proposed algorithm is significantly better than that of URA schemes designed in \cite{unsourced} and \cite{newT}. There are two main reasons, the first is that the proposed algorithm exploits the prior statistical distributions of the sparse channel, which is ignored by the algorithms in \cite{unsourced} and \cite{newT}. On the other hand, the RIS can reflect impinging
electromagnetic waves with a controllable phase shift via the
help of a smart controller. Through the passive beamforming designed in Section III-A, the RIS is able to provide an extra high-quality channel link to overcome the unfavourable propagation conditions of URA systems.

\section{Conclusion}
In this paper, we proposed a new RIS-aided URA architecture and formulated joint active device separation and channel estimation as a coupled high-order tensor problem. To solve the problem, a novel probabilistic modeling and an automatic learning algorithm were proposed under the framework of Bayesian inference. It is found that the deployment of a RIS can effectively improve the performance of URA in 6G wireless networks.

\begin{appendices}
\section{Lemma 1 and Lemma2}
In this section, we provide some results on normal distribution and $\mathbf{G}$, which will be used for the derivation of individual variables in Section IV.

\emph{Lemma 1}: Given the vector-valued normal distribution $\mathcal{CN}_{NK}(\mathbf{x}|\mathbf{a}_i,\mathbf{A}_i)$ with the mean vector $\mathbf{a}_i$ and covariance $\mathbf{A}_i$ for $i=1,\cdots,n$, we have
\begin{align}\label{gaum}
&\!\!\mathop \prod \limits_{i=1}^{n}\mathcal{CN}_{NK}(\mathbf{x}|\mathbf{a}_i,\mathbf{A}_i)
\propto \exp \left[-\frac{1}{2}\mathbf{x}^H\left(\mathop \sum \limits_{i=1}^{n}\mathbf{A}_i^{-1}\right)\mathbf{x}\!-\!\mathbf{x}^H\right.\nonumber\\
&\!\!\left.\times\left(\!\mathop \sum_{i=1}^{n}\mathbf{A}_i^{-1}\mathbf{a}_i\!\right)\!-\!\left(\!\mathop \sum_{i=1}^{n}\mathbf{A}_i^{-1}\mathbf{a}_i\!\right)^H\mathbf{x}\right] \!\!= \!\mathcal{CN}_{NK}(\mathbf{x}|\mathbf{c},\mathbf{C}),
\end{align}
with
$
\mathbf{C}=\left(\mathop \sum \limits_{i=1}^{n}\mathbf{A}_i^{-1}\right)^{-1}$ and $\mathbf{c}=\mathbf{C}\left(\mathop \sum_{i=1}^{n}\mathbf{A}_i^{-1}\mathbf{a}_i\right)$.

\emph{Lemma 2}: If $\mathrm{vec}(\mathbf{G}^H)$ obeys the vector-valued normal distribution $\mathcal{CN}_{NK}(\mathrm{vec}(\mathbf{G}^H)|\mathbf{u},\boldsymbol{\Omega})$, then $\mathbf{G}$ is said to follow the matrix-variate normal distribution $\mathcal{CN}_{N\times K}(\mathbf{G}|\mathbf{M}_G,\boldsymbol{\Omega})$ with mean matrix $\mathbf{M}_G$  and covariance matrix $\boldsymbol{\Omega}$. Herein, the mean matrix $\mathbf{M}_G\in \mathbb{C}^{N \times K}$ is obtained by rearranging vector $\mathbf{u}$.
This further implies that
\begin{align}\label{mean}
\mathbb{E}[\mathbf{G}^H\mathbf{G}]&=\mathbf{M}_G^H\mathbf{M}_G+\mathop\sum \limits_{n=1}^{N}\boldsymbol{\Omega}_{n,n}^b,\\
\mathbb{E}[\mathbf{G}^H\mathbf{P}_l^H\mathbf{P}_l\mathbf{G}]&=\mathbf{M}_G^H\mathbf{P}_l^H\mathbf{P}_l\mathbf{M}_G+
\mathop\sum \limits_{i=1}^{N}\mathop\sum \limits_{j=1}^{N}\{\mathbf{P}_l^H\mathbf{P}_l\}(i,j)\boldsymbol{\Omega}_{i,j}^{b},\label{mean1}
\end{align}
where $\boldsymbol{\Omega}_{i,j}^{b}\in\mathbb{C}^{K\times K}$ is the $(i,j)$th block of $\boldsymbol{\Omega}$.
\end{appendices}


\begin{thebibliography}{1}
\bibitem{GC}
X. Shao, L. Cheng, X. Chen, C. Huang, and D. W. K. Ng, ``A Bayesian tensor approach to enable RIS for 6G massive unsourced random access," \emph{IEEE Global Commun. Conf. (GLOBECOM)}, Madrid, Spain, Dec. 2021.

\bibitem{coop}
H. Xiao, W. Chen, J. Fang, B. Ai, and I. J. Wassell, ``A grant-free method for massive machine-type communication with backward activity level estimation," \emph{IEEE Trans. Signal Process.}, vol. 68, pp. 6665-6680, Dec. 2020.

\bibitem{6G}
X. Shao, X. Chen, D. W. K. Ng, C. Zhong, and Z. Zhang, ``Cooperative active device separation: Sourced and unsourced massive random access paradigms," \emph{IEEE Trans. Signal Process.}, vol. 68, pp. 6578-6593, Dec. 2020.
%

\bibitem{ampgao}
M. Ke, Z. Gao, Y. Wu, X. Gao, and R. Schober, ``Compressive sensing-based adaptive active user detection and channel estimation: Massive access meets massive MIMO," \emph{IEEE Trans. Signal Process.}, vol. 68, pp. 764-779, Jan. 2020.

\bibitem{shaoiot}
X. Shao, X. Chen, C. Zhong, J. Zhao, and Z. Zhang, ``A Unified design of massive access for cellular Internet of Things," \emph{IEEE Intern. Things J.}, vol. 6, no. 2, pp. 3934-3947, 2019.

\bibitem{PersUra}
Y. Polyanskiy, ``A perspective on massive random-access," in \emph{Proc. IEEE Int. Symp. Inf. Theory}, Aachen, Germany, Jun. 2017, pp. 2523-2527.

\bibitem{bayf}
A. Fengler, S. Haghighatshoar, P. Jung, and G. Caire, ``Non-Bayesian activity detection, large-scale fading coefficient estimation, and unsourced random access with a massive MIMO receiver," \emph{IEEE Trans. Inform. Theory}, vol. 67, no. 5, pp. 2925-2951, May 2021.

\bibitem{unsourced}
A. Fengler, G. Caire, P. Jung, and S. Haghighatshoar, ``Massive MIMO unsourced random access," Jan. 2019, [Online]. Available: http://arxiv.org/abs/1901.00828.

\bibitem{wuyongUra}
X. Xie, Y. Wu, J. Gao, and W. Zhang, ``Massive unsourced random access for massive MIMO correlated channels", \emph{IEEE Global Commun. Conf. (GLOBECOM)}, Taipei, Taiwan, Dec. 2021.

\bibitem{shi}
S. Xia and Y. Shi, ``Intelligent reflecting surface for massive device connectivity: Joint active device separation and channel estimation," \emph{IEEE Intern. Conf. Acoustics, Speech and Signal Process. (ICASSP)}, Barcelona, Spain, 2020, pp. 5175-5179.

\bibitem{shaommv}
X. Shao, X. Chen, C. Zhong, and Z. Zhang, ``Exploiting simultaneous low-rank and sparsity in delay-angular domain for Millimeter-Wave/Terahertz wideband massive access," \emph{IEEE Trans. Wireless Commun.}, vol. 21, no. 4, pp. 2336-2351, Apr. 2022.

\bibitem{kwanro}
X. Yu, D. Xu, Y. Sun, D. W. K. Ng, and R. Schober, ``Robust and secure wireless communications via intelligent reflecting surfaces," \emph{IEEE J. Sel. Areas Commun.}, vol. 38, no. 11, pp. 2637-2652, Nov. 2020.

\bibitem{rui}
Q. Wu, S. Zhang, B. Zheng, C. You, and R. Zhang, ``Intelligent reflecting surface aided wireless communications: A tutorial," \emph{IEEE Trans. Commun.}, vol. 69, no. 5, pp. 3313-3351, May 2021.



\bibitem{pan}
C. Huang, S. Hu, G. C. Alexandropoulos, A. Zappone, C. Yuen, R. Zhang, R. D. Renzo, and M. Debbah, ``Holographic MIMO surfaces for 6G wireless networks: Opportunities, challenges, and trends," \emph{IEEE Wireless Commun.}, vol. 27, no. 5, pp. 118-125, Oct. 2020.

\bibitem{imag}
J. Yao, Z. Zhang, X. Shao, \emph{et al}, ``Concentrative intelligent reflecting surface aided computational imaging via fast block sparse Bayesian learning," in \emph{Proc. IEEE Veh. Technol. Conf. (VTC)}, Jun. 2021, pp. 1-6.



\bibitem{huang}
C. Huang, A. Zappone, G. C. Alexandropoulos, M. Debbah, and C. Yuen, ``Reconfigurable intelligent surfaces for energy efficiency in wireless communication," \emph{IEEE Trans. Wireless Commun.}, vol. 18, no. 8, pp. 4157-4170, Aug. 2019.

\bibitem{irssensing}
X. Shao, C. You, W. Ma, X. Chen, and R. Zhang, ``Target sensing with intelligent reflecting surface: Architecture and performance," \emph{IEEE J. Sel. Areas Commun.}, vol. PP, no, 99, pp. 1-1, Mar. 2022.



\bibitem{RISsparse}
H. Liu, X. Yuan, and Y. -J. A. Zhang, ``Matrix-calibration-based cascaded channel estimation for reconfigurable intelligent surface aided multiuser MIMO," \emph{IEEE J. Sel. Areas Commun.,} vol. 38, no. 11, pp. 2621-2636, Nov. 2020.

\bibitem{cou}
M. S. Rensen and L. De Lathauwer, ``Coupled canonical polyadic decompositions and (coupled) decompositions in multilinear rank-($L_{r,n},L_{r,n},1$) terms-part I: Uniqueness" \emph{SIAM J. Matrix Analysis and Appli.}, vol. 36, no. 2, pp. 496-522, 2015.

\bibitem{cou2}
L. De Lathauwer and E. Kofidis, ``Coupled matrix-tensor factorizations-the case of partially shared factors" \emph{Asilomar Conference on Signals, Systems, and Computers.} 2017, pp. 711-715.

\bibitem{tenrank}
T. G. Kolda and B. W. Bader, ``Tensor decompositions and applications," \emph{SIAM Rev.}, vol. 51, pp. 455-500, 2009.

\bibitem{chengpro}
L. Cheng, Y.-C. Wu, and H. V. Poor, ``Probabilistic tensor canonical polyadic decomposition with orthogonal factors," \emph{IEEE Trans. Signal Process.,} vol. 65, no. 3, pp. 663-676, Feb., 2017.

\bibitem{zhao}
Q. Zhao, L. Zhang, and A. Cichocki, ``Bayesian CP actorization of incomplete tensors with automatic rank determination" \emph{IEEE Trans. Pattern Analy. Machine Intell.}, vol. 37, no. 9, pp. 1751-1763, Sept. 2015.

\bibitem{cheng1}
L. Cheng and Q. Shi, ``Towards overfitting avoidance: Tuning-free tensor-aided multi-user channel estimation for 3D massive MIMO communications," \emph{IEEE J. Sel. Topics Signal Process.}, vol. 15, no. 3, pp. 832-846, Apr. 2021.

\bibitem{cheng2}
L. Cheng, X. Tong, S. Wang, Y. Wu, and H. V. Poor, ``Learning nonnegative factors from Tensor data: Probabilistic modeling and inference algorithm," \emph{IEEE Trans. Signal Process.}, vol. 68, pp. 1792-1806, Feb. 2020.


\bibitem{power}
W. U. Bajwa, J. Haupt, A. M. Sayeed, and R. Nowak, ``Compressed channel sensing: A new approach to estimating sparse multipath channels," \emph{Proc. IEEE}, vol. 98, no. 6, pp. 1058-1076, Jun. 2010.

\bibitem{lee}
A. L. Swindlehurst, E. Ayanoglu, P. Heydari, and F. Capolino, ``Millimeter-wave massive MIMO: The next wireless revolution?" \emph{IEEE Commun. Mag.}, vol. 52, no. 9, pp. 56-62, Sep. 2014.

\bibitem{shaodim}
X. Shao, X. Chen, and R. Jia, ``A dimension reduction-based joint activity detection and channel estimation algorithm for massive access," \emph{IEEE Trans. Signal Process.}, vol. 68, pp. 420-435, Jan. 2020.



\bibitem{beixiong}
B. Zheng, C. You, W. Mei, and R. Zhang, ``A survey on channel estimation and practical passive beamforming design for intelligent reflecting surface aided wireless communications," \emph{IEEE Commun. Surv. Tuts.}, vol. 24, no. 2, pp. 1035-1071, Jun. 2022.

\bibitem{sta}
C. Hu, L. Dai, S. Han, and X. Wang, ``Two-timescale channel estimation for reconfigurable intelligent surface aided wireless communications," \emph{IEEE Trans. Commun.}, vol. 69, no. 11, pp. 7736-7747, Nov. 2021.

\bibitem{xu}
X. Yu, V. Jamali, D. Xu, D. W. K. Ng, and R. Schober, ``Smart and reconfigurable wireless communications: From IRS modeling to algorithm design," \emph{IEEE Wireless Commun.}, vol. 28, no. 6, pp. 118-125, Dec. 2021.

\bibitem{co}
F. Laue, V. Jamali, and R. Schober, ``IRS-assisted active device separation," \emph{International Workshop Signal Process. Advances Wireless Commun. (SPAWC)}, July 2022.

\bibitem{emlihard}
E. Bj\"{o}rnson and L. Sanguinetti, ``Rayleigh fading modeling and channel hardening for reconfigurable intelligent surfaces," \emph{IEEE Wireless Commun. Lett.}, vol. 10, no. 4, pp. 830-834, Apr. 2021.

\bibitem{convex}
S. Boyd and L. Vandenberghe, \emph{Convex Optimization}, New York, NY, USA: Cambridge University Press, 2004.

\bibitem{mosek}
Z.-Q. Luo, W.-K. Ma, A. M. So, Y. Ye, and S. Zhang, ``Semidefinite relaxation of quadratic optimization problems," \emph{IEEE Signal Process. Mag.}, vol. 27, pp. 20-34, May 2010.

\bibitem{newT}
A. Decurninge, I. Land, and M. Guillaud, ``Tensor-based modulation for unsourced massive random access," \emph{IEEE Wireless Commun. Let.}, vol. 10, no. 3, pp. 552-556, 2020.


\bibitem{blind1}
L. Chen and X. Yuan, ``Blind multiuser detection in massive MIMO channels with clustered sparsity," \emph{IEEE Wireless Commun. Lett.}, vol. 8, no. 4, pp. 1052-1055, Aug. 2019.

\bibitem{ste}
S. M. Kay, \emph{Fundamentals of Statistical Signal Srocessing}, Prentice Hall PTR, 1993.

\bibitem{vi}
M. J. Wainwright and M. I. Jordan, ``Graphical models, exponential families, and variational inference," \emph{Found. Trends Mach. Learn.,} vol. 1, no. 102, pp. 1-305, Jan. 2008.

\bibitem{grcode2}
K. Ngo, A. Decurninge, M. Guillaud, and S. Yang, ``Cube-Split: A structured Grassmannian constellation for non-coherent SIMO communications," \emph{IEEE Trans. Wireless Commun.,} vol. 19, no. 3, pp. 1948-1964, Mar. 2020.

\end{thebibliography}
\end{document}